\documentclass[aps,pra,floatfix,twocolumn,showpacs,tightenlines,groupedaddress,superscriptaddress,amsmath]{revtex4}
\usepackage{verbatim}

\ifx\pdftexversion\undefined
  \usepackage[dvips]{color,graphicx}
\else
  \usepackage[pdftex]{color,graphicx}
\fi

\def \ra{\rightarrow}

\def \Bdpa{\Omega_B}

\def \hH{\hat{H}}
\def \hI{\hat{I}}

\def \nbar{\bar{n}}
\def \bS{\bar{S}}

\def \tomega{\tilde{\omega}}

\def\nonlinearconst{\Lambda}

\def\tlam{\tilde{g}}

\def\cavityavg{\sqrt{\bar{n}}}
\def\photonavg{\bar{n}}

\def \be{\begin{equation}}
\def \ee{\end{equation}}

\def \ba{\begin{eqnarray}}
\def \ea{\end{eqnarray}}

\def \skap{\sqrt{\kappa}}

\def \hP{\hat{P}}

\def \hz{\hat{z}}

\def\hadag{{\hat a}^\dagger}

\def\ha{\hat{a}}
\def \hbin{\hat{b}_{\rm in}}

\def \hbout{\hat{b}_{\rm out}}
\def \hd{\hat{d}}

\def\hddag{{\hat d}^\dagger}
\def \hxi{\hat{\xi}}

\def \hF{\hat{F}}

\def\tdel{\widetilde{\Delta}}
\def\hX{\hat{X}}
\def\hP{\hat{P}}

\def\T{\mathbf{T}}

\def\M{\mathbf{M}}

\def	\bse{\begin{subequations}}
\def	\ese{\end{subequations}}

\begin{document}

\title{Quantum limited amplification with a nonlinear cavity detector}
\author{C. Laflamme}
\affiliation{Physics Department, McGill University, Montreal,
Quebec, Canada H3A 2T8}
\author{A. A. Clerk}
\affiliation{Physics Department, McGill University, Montreal,
Quebec, Canada H3A 2T8}

\date{Nov. 25, 2010}

\begin{abstract}
We consider the quantum measurement properties of a driven cavity with a Kerr-type nonlinearity which is used to amplify a dispersively coupled input signal.  Focusing on an operating regime which is near a bifurcation point, we derive simple asymptotic expressions describing the cavity's noise and response.  We show that the cavity's backaction and imprecision noise allow for quantum limited linear amplification and position detection only if one is able to utilize the sizeable correlations between these quantities.  This is possible when one amplifies a non-resonant signal, but is not possible in QND qubit detection.  We also consider the possibility of using the nonlinear cavity's backaction for cooling a mechanical mode.
\end{abstract}

\maketitle



\section{Introduction}
\label{sec:nonlinear}

A number of recent experiments have made use of driven microwave transmission-line resonators for sensitive, near-quantum limited measurements.  These include measurements of the position of a nanomechanical oscillator near the standard quantum limit \cite{Lehnert09, Schwab09}, as well as measurements of single and multi-qubit systems in circuit QED setups \cite{Schuster05, Majer07}.  Such experiments use the microwave cavity as an ``op-amp" type amplifier \cite{ClerkRMP}, where the signal to be detected (e.g.~the position $x$ of a mechanical resonator or the $\sigma_z$ operator of a qubit) is dispersively coupled to the microwave cavity, meaning that the cavity frequency depends on the signal.  When the cavity is driven, the resulting modulation of the cavity frequency by the signal leads to a modulation of the phase of the reflected beam from the cavity.   By monitoring this phase (e.g.~via homodyne interferometry), one has essentially amplified the signal.

Most experiments using a cavity for such dispersive measurements and amplification have not exploited nonlinearities in the cavity-- the cavity is just a driven, damped harmonic oscillator.  The resulting measurement and amplification properties of the system are well understood.  In particular, it is known that this system can be used for quantum-limited linear amplification, meaning that the total added noise of the measurement can be as small as allowed by quantum mechanics (see, e.g.,~Ref.~\cite{ClerkRMP} for a pedagogical discussion). 

While the linear-cavity regime is certainly useful, it is also interesting to consider another possibility afforded by microwave circuit cavities:  they can be engineered to have strong Kerr-type nonlinearities through the use of Josephson junctions \cite{Yurke87, Yurke89, Blencowe08}. 
The resulting nonlinear cavity can then be used for amplification in ways not possible with a linear cavity.  Attention has largely focused on using such devices in the ``scattering" mode of operation, where the signal to be amplified is incident on the cavity from a coupled transmission line, and where backaction effects are irrelevant.  Experiments using this mode have realized single-quadrature amplification and squeezing \cite{Yurke89, Yurke90, Lehnert08, Nakamura08}; this operation mode has also been the subject of many theoretical treatments, e.g. Refs.~\cite{Yurke87, Yurke89, Yurke06, Milburn08}.
 For qubit detection, another possibility is to use the bifurcation in a nonlinear cavity to give a latching-type measurement, where the final dynamical state of the cavity depends on the initial state of the qubit \cite{Siddiqi04, Lupascu06}; this scheme has also received theoretical attention \cite{Wilhelm10}.

In this work, we will instead study theoretically the quantum measurement properties of a driven nonlinear cavity in the operation mode most relevant to experiments in nanomechanics and quantum information, the so-called ``op-amp" mode of operation described above.  Unlike the scattering mode studied in \cite{Yurke87, Yurke89, Yurke06}, here backaction is indeed relevant, and plays a crucial role in enforcing the quantum limit on the added noise:  
to reach the quantum limit, the backaction noise must be as small as allowed by quantum mechanics \cite{Braginsky92, ClerkRMP}.  We will focus exclusively on regimes where there is no multistability in the cavity dynamics (in contrast to the bifurcation amplifier setup).
We note that experiments using a nonlinear microwave cavity amplifier in the ``op-amp" mode discussed here have recently been performed.  Vijay et al. have constructed a nonlinear cavity formed by a SQUID \cite{Vijay10}, and have used this to detect a dispersively coupled superconducting qubit \cite{Vijay10b}.  A recent experiment by Ong et al.~\cite{Esteve10} uses a nonlinear microwave formed from a transmission line resonator and a Josephson junction to detect a dispersively-coupled qubit; this experiment also investigated backaction effects.

\begin{figure}[t!]
	 \centering
	\includegraphics[width=0.95 \columnwidth]{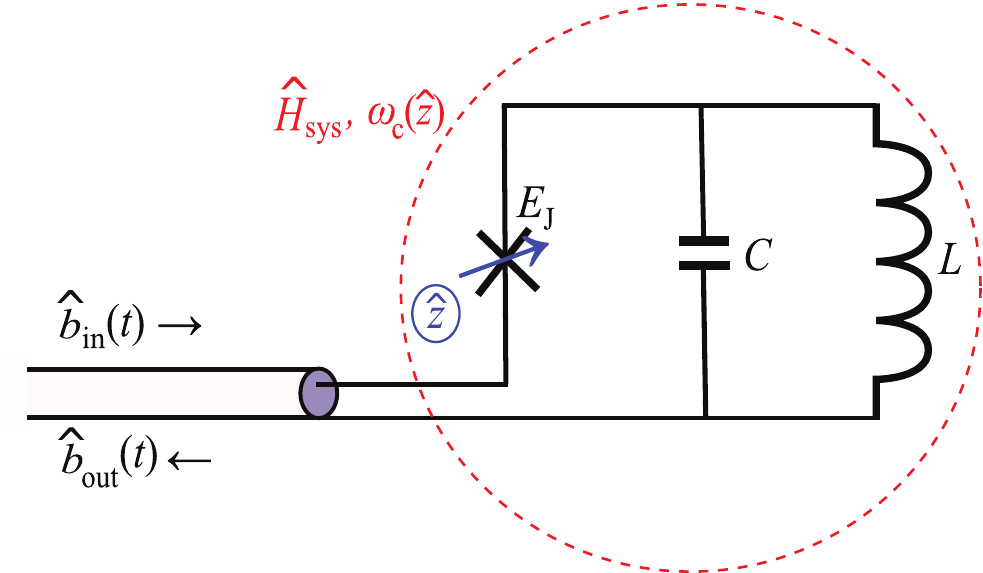}
	\caption{
		Schematic figure of a realization of the nonlinear cavity amplifier.  The cavity is formed by an $LC$ circuit containing a 
		Josephson junction	(energy $E_J$).  The cavity is damped and driven by a coupled transmission line; $\hbin$ and $\hbout$  denote the
		input and output fields in the transmission. The input signal $\hz$ is a flux which controls the value of $E_J$, 
		and hence the frequency of the cavity $\omega_c$.  An experimental realization of this system is presented in Ref.~\cite{Vijay10}.}
\label{fig:Schematic}
\end{figure}

Our analysis focuses on operation points close (but not past) the bifurcation in the cavity response, a regime which yields extremely large small-signal, low-frequency amplification gain.  The approach we use is standard:  we linearize the cavity dynamics about its mean, classical value, and use the resulting linear quantum Langevin equations to study the noise properties of the cavity detector.  This allows us to asses its ability to reach the ``op-amp" amplifier quantum limit; a related analysis is presented in Ref.~\cite{Blencowe08}.  Despite this standard approach, we find a number of surprising conclusions that seem not to have been appreciated in the existing literature.  In particular, we show that the nonlinear cavity near the bifurcation is equivalent to a degenerate parametric amplifier (DPA) driven with a {\it detuned pump} \cite{Carmichael84}.  The value of this effective detuning is not an independent parameter, and tends to a universal value as one approaches the bifurcation.  This mapping allows us to derive simple, analytic asymptotic expressions for the cavity's noise and gain that are universally valid as one approaches the bifurcation point.  

In the low frequency limit, large-gain limit, we find that the imprecision noise of the cavity is precisely four times what would be expected of an ideal, resonantly-pumped degenerate parametric amplifier with equivalent gain (c.f.~Eqs.~(\ref{eqs:IdealDPA}),(\ref{eq:szz_ideal})).  We also show, somewhat surprisingly, that the cavity's backaction noise at low frequencies is always given by the same, simple expression valid for a linear cavity, an expression which is usually interpreted as describing the overlap of displaced coherent states (c.f. Eqs.~(\ref{eq:SFFOverlap}),(\ref{eq:SFFApprox})).  We find that the nonlinear cavity amplifier is quantum limited at low frequencies, but {\it only} if one can make use of the large correlations between the backaction and imprecision noises.  Such correlations cannot be utilized simply in QND qubit detection; hence, the cavity amplifier misses the quantum limit on QND qubit detection by a large factor.  

We also use our approach to study the possibility of using the backaction of a nonlinear driven microwave cavity to cool a nanomechanical resonator.  Near the bifurcation point, we find an extremely simple expression for the effective temperature of the nonlinear cavity's backaction (c.f.~Eq.~(\ref{eq:occ_number})); surprisingly, the only relevant cavity parameter is its damping rate $\kappa$.  We also show that a driven nonlinear cavity is far better at cooling a low frequency mechanical oscillator than a corresponding driven linear cavity with comparable parameters.  This latter conclusion matches what was found by Nation et al. \cite{Blencowe08},  who studied backaction cooling by a noninear cavity numerically over a wide range of cavity parameters, including regimes where there is bistability in the cavity dynamics.  Aspects of cooling and heating using a driven nonlinear cavity were also addressed by Dykman \cite{Dykman78}.

The remainder of this paper is organized as follows.    In Sec.~\ref{sec:Basics}, we review the basics of how one uses a nonlinear cavity as a linear ``op-amp" style amplifier, and review the formulation and origin of the quantum limit applicable here.  In Sec.~\ref{sec:Mapping}, we show how near the bifurcation, the nonlinear cavity is equivalent to a DPA driven by a detuned pump.  In Sec.~\ref{sec:LowFrequencyAnalysis}, we use this mapping to derive asymptotic expressions for the cavity's noise and amplifier gain near the bifurcation point, and assess its ability to reach the quantum limit for small signal frequencies.  Sec.~\ref{sec:FiniteFrequencyAnalysis} extends this analysis to non-zero signal frequencies.  Finaly, in Sec.~\ref{sec:Cooling}, we consider the asymmetric quantum backaction noise of the cavity, and assess the possibility of using the cavity for backaction cooling a mechanical oscillator.

\section{Basics of a nonlinear cavity amplifier}   
\label{sec:Basics}

\subsection{System Hamiltonian}

The Hamiltonian of a cavity detector with a Kerr-type nonlinearity has the general form
\be
	\label{eq:ham_cavity}
 	\hH = \hH_{\rm sys} + \hH_{\kappa} 
	=   \hbar \omega_c \hadag \ha - \hbar \nonlinearconst \hadag \hadag \ha \ha + \hH_{\kappa},
\ee
where  $\omega_c$ is the cavity resonance frequency, $\nonlinearconst$ is the Kerr constant.  We take $\nonlinearconst > 0$ in what follows, 
as is appropriate for a microwave cavity incorporating Josesphson junctions; the results are easily generalized to $\nonlinearconst < 0$.
The term $\hat{H}_\kappa$ represents the damping (at rate $\kappa$) and driving of the cavity due to its coupling to extra cavity modes (e.g.~in a microwave circuit, to the transmission line used 
to drive the cavity).  Derivations of this Hamiltonian for microwave circuits incorporating Josephson junctions are presented in many places in the literature, and we do not repeat them here (see~e.g.~Refs.~\cite{Yurke87, Yurke89, Yurke06, Blencowe08, Vijay10});
a schematic is presented in Fig.~\ref{fig:Schematic}.  In writing Eq.~(\ref{eq:ham_cavity}) we have assumed the relevant case of a high-$Q$ cavity and thus made use of the rotating wave approximation to write the nonlinear term.  We will also be interested throughout in the case of a weak nonlinearity, $\nonlinearconst \ll \kappa $.   For clarity, we focus exclusively on the ideal case where there is no internal cavity loss; we also focus on the case of a one-sided cavity.  Our analysis could be easily generalized to incorporate either a two-sided cavity or internal loss (see e.g.~Ref.~\cite{Yurke06}).

\begin{figure}[t!]
	\centering
	\includegraphics[width= 0.91 \columnwidth]{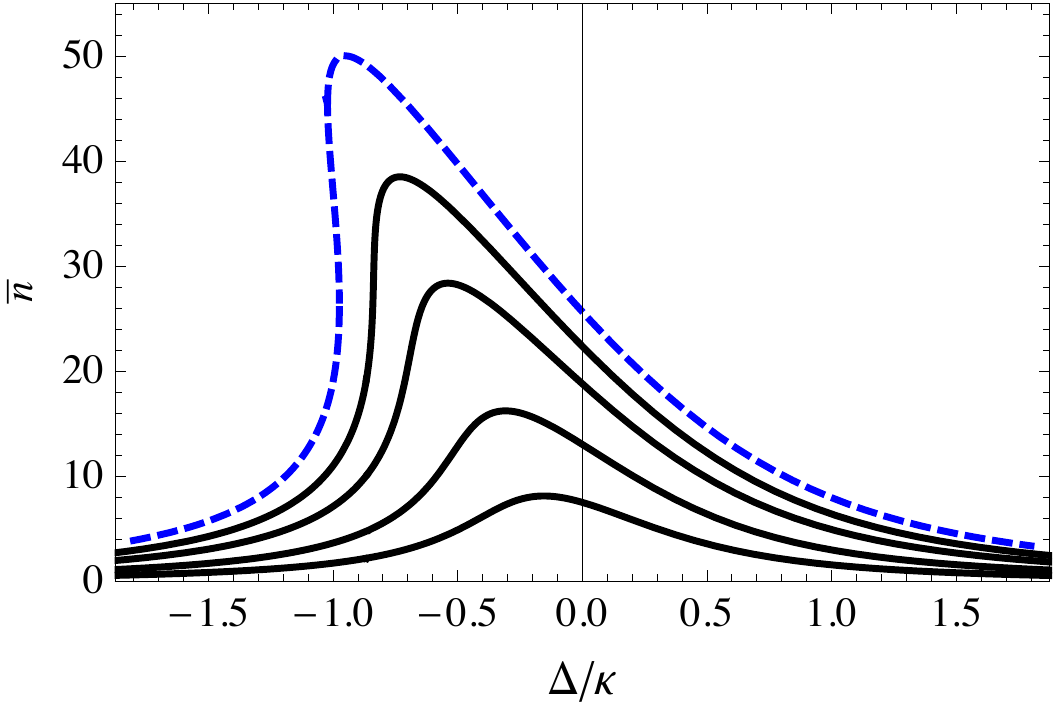}
	\caption{
		Average cavity photon number $\bar{n}$ versus drive detuning $\Delta = \omega_d - \omega_c$, 
		for various driving strengths  $\bar{b}_{\rm in}$.  
		The curves show the evolution of the cavity response as one goes through the bifurcation:  
		solid curves are for $\bar{b}_{\rm in} < \bar{b}_{\rm in,bif}$, while the dashed curve is for  	 
		$\bar{b}_{\rm in} > \bar{b}_{\rm in,bif}$.  The diverging slope of $d \bar{n} / d \Delta$ near the bifurcation allows for amplification with a large gain.	}
	\label{fig:NBar}
\end{figure}

Unlike a linear cavity, the nonlinear cavity described by Eq.~(\ref{eq:ham_cavity}) can undergo a bifurcation as a function of its parameters from a regime where the average cavity photon number $\nbar = \langle \ha^\dag \ha \rangle$ is a single valued function of the 
drive frequency $\omega_d$, to a regime where it is multivalued.  For drive strengths just below the bifurcation threshold, $\nbar$
is a single-valued function of drive frequency, but exhibits a very pronounced slope (see Fig.~\ref{fig:NBar}).  
This extreme sensitivity to cavity frequency makes the cavity an extremely sensitive dispersive detector and amplifier in this regime.  However, it is not {\it a priori} obvious whether the cavity's noise in this regime is small enough to allow quantum limited performance.  Answering this question is our main goal.  

\subsection{Quantum limit on amplification in the op-amp mode of operation}

We focus throughout on the op-amp mode of amplifier operation, where the input signal to be detected (described by an operator $\hz$) is coupled directly to the cavity photon number:
\begin{eqnarray}
	H_{\rm int} = A  \hadag \ha \cdot \hz  \equiv \hF \cdot \hz.
	\label{eq:Hint}
\end{eqnarray}
The operator $\hz$ could represent (for example) the position of a nanomechanical beam (as considered in Ref.~\cite{Blencowe08}) or the signal flux applied to a SQUID circuit (as in Ref.~\cite{Vijay10}).  As a result of this dispersive coupling, the cavity frequency and reflected-beam phase shift become $z$-dependent;  one can thus amplify $z(t)$ by monitoring this phase. We will focus here on homodyne detection, where the output beam is interfered with a classical reference beam, and the resulting intensity measured using a square-law detector.  Letting $\hbout(t)$ denote the output field from the cavity
(as defined in standard input-output theory \cite{Gardiner85, Gardiner00}), the measured homodyne intensity will be described by an operator $\hat{I}(t)$
\begin{eqnarray}
	\hat{I}  = B \sqrt{\kappa/2} \left( e^{\i \phi} \hbout(t) + \textrm{ h.c.} \right),
	\label{eq:IDef}
\end{eqnarray}
where $\phi$ is the phase of the classical reference beam, and $B$ is a dimensionless constant proportional to the amplitude of this beam.  Note that $\hI$ has units corresponding to a photon flux.  As the value of $B$ plays no role in what follows (it is just a scale factor for the output), we set $B=1$ without loss of generality in what follows.

We will be interested in weak enough couplings that our cavity acts as a linear amplifier.  As such, we have a linear relation between the input signal and cavity output:
\begin{eqnarray}
	\langle \hI(t) \rangle & = &
		 \int_{-\infty}^{\infty} dt' \chi_{IF}(t-t') \langle \hz(t') \rangle,
	\label{eq:ChiIFDefn}
\end{eqnarray}
where $\chi_{IF}(t) \propto A$ is the forward gain of the amplifier, and is determined by a standard Kubo formula \cite{ClerkRMP}.

The amplifier output (i.e.~$I$) will have fluctuations even in the absence of any coupling to the detector; these are described by the symmetrized spectral density:
\begin{eqnarray}
	\bS_{II}[\omega] = \frac{1}{2} \int_{-\infty}^{\infty} dt e^{i \omega t} 
		\left\langle \{ \hI(t), \hI(0) \} \right\rangle.
	\label{eq:SIDefn}
\end{eqnarray}
Again, as we are interested in linear amplification, the expectation value is taken in with respect to the state of the {\it uncoupled} detector (i.e.~$A=0$).  It is useful and standard to think of these intrinsic output fluctuations in terms of effective signal (i.e.~$z$) fluctuations; we thus introduce the imprecision noise spectral density:
\ba
	\bS_{zz}[\omega] = \bS_{II}[\omega]  / \left| \chi_{IF}[\omega] \right|^2.
	\label{eq:ImprecisionNoiseDefn}
\ea

In the op-amp mode of operation, a second crucial aspect of the amplifier's noise is its backaction.  By virtue of the the detector-signal coupling in Eq.~(\ref{eq:Hint}), the operator $\hF = A \hadag \ha$ (i.e.~the cavity photon number) acts as a noisy backaction force on the signal.  Extra fluctuations in $\hz$ due to this stochastic force will necessarily increase the noise in the output of the amplifier, and are thus part of the total added noise of the amplifier.  
The backaction force noise is characterized by a symmetrized noise spectral density $\bS_{FF}[\omega]$ defined analogously to Eq.~(\ref{eq:SIDefn}).

We thus have that the total amplifier contribution to the output noise has contributions from both imprecision and backaction noise.  It is convenient (and common) to think of this total added noise in terms of a noise temperature $T_N[\omega]$:  the total amplifier added noise at frequency $\omega$ is equivalent to the extra equilibrium noise we would get by raising the temperature of the signal source by $T_N[\omega]$
\footnote{We use the standard convention in which the noise temperature $T_N[\omega]$ is defined by assuming the signal source is initially at a temperature much larger than $\hbar \omega$  This definition leads to the standard bound given in Eq.~(\ref{eq:TNSQL}).}.  
This quantity is relevant no matter what the signal, be it the position of a harmonic oscillator or the voltage produced by some input circuit;
we also stress that achieving the quantum limit on the noise temperature is equivalent to achieving the quantum limit on continuous weak displacement detection \cite{ClerkRMP}.

Minimizing the noise temperature at a given frequency requires one to first optimize the signal source's susceptibility $\chi_{zz}[\omega]$.  This linear-response susceptibility tells us how the average value of $\hz$ changes in response to a perturbation which couples to $\hz$, i.e.:
\begin{eqnarray}
	\delta \langle \hz(t) \rangle = \int_{-\infty}^{\infty} dt' \chi_{zz}(t-t') \langle \hF(t') \rangle.
\end{eqnarray}
Optimizing the total added noise over the coupling strength and phase of the signal source's susceptibility 
$\chi_{zz}[\omega]$ yields a standard bound on $T_N[\omega]$ \cite{ClerkRMP}:
\begin{eqnarray}
	\frac{k_{\rm B} T_{N}[\omega]}{\hbar \omega} & \geq &
	\frac{1}{\hbar} \Bigg(
		 \sqrt{
		\bS_{zz}[\omega] \bS_{FF}[\omega] - 
			\left[ \mathrm{Re }\left(\bS_{zF}[\omega] \right) \right]^2
		} \nonumber \\
	&&
	- {\rm Im  }\bS_{zF}[\omega]
	\Bigg),
	\label{eq:TNDefn}
\end{eqnarray}
where the inequality becomes an equality for an optimal source susceptibility satisfying:
\bse
\label{eqs:OptimalChi}
\begin{eqnarray}
	| \chi_{zz}[\omega] | & = & 
		\sqrt{ \bS_{zz}[\omega] /  \bS_{FF}[\omega]  }
		\label{eq:OptimalChiMag}\\
	\frac{ \textrm{Re } \chi_{zz}[\omega] }{ | \chi_{zz}[\omega] |}
		&  = &   
			\frac{ \textrm{Re } \bS_{zF}[\omega]}{ \sqrt{  \bS_{zz}[\omega] \bS_{FF}[\omega] } } .
		\label{eq:OptimalChiPhase}
\end{eqnarray}
\ese
We have introduced the correlator $\bS_{zF}$ which describes possible correlations between backaction and imprecision noises:
\ba
	\bS_{zF}[\omega] & = & \frac{  \bS_{IF}[\omega]  }{  \chi_{IF}[\omega]}
	= 
	 \frac{ 
	 	\int_{-\infty}^{\infty} dt e^{i \omega t} \langle \{ \hI(t), \hF(0) \} \rangle 
	 }{2  \chi_{IF}[\omega]}.	
\ea

Consider the simple case where the signal frequency $\omega$ is much smaller than the relevant frequency scales of the cavity; we may thus focus on the noise temperature in the $\omega \ra 0$ limit.  Using the fact that there cannot be any out-of-phase noise correlations at zero frequency (i.e.~$\bS_{zF}[0] = \textrm{Re }\bS_{zF}[0] $), the zero-frequency form of the fundamental Heisenberg inequality on detector noise \cite{ClerkRMP}:
\ba
	\bS_{zz}[0] \bS_{FF}[0] - \left( \bS_{zF}[0] \right)^2 \geq \hbar^2/4,
	\label{eq:QNConstraintZ}
\ea
implies that 
\be
	k_{\rm B}T_N \geq \hbar \omega / 2,
	\label{eq:TNSQL}
\ee
i.e.~the added noise amplifier must at least as large as the zero-point noise of the signal source
\cite{Braginsky92, ClerkRMP}
\endnote{Note that we have used the form of the quantum noise inequality corresponding to a vanishing ``reverse gain", i.e.~coupling to $\hI$ cannot change the average value of $\hF$.  The vanishing of the reverse gain for our nonlinear cavity amplifier is explicitly demonstrated in 
Appendix~\ref{sec:appendix_dpa}.}.  
We stress that while the conclusion may appear similar, the ``op-amp" quantum limit considered here is {\it not} identical to the quantum limit  on the ``scattering" mode described in the seminal works by Haus and Mullen \cite{Haus62} and Caves \cite{Caves82}:  the scattering-mode quantum limit does not involve backaction.  Moreover, an amplifier may reach the quantum limit in the scattering mode but not in the op-amp mode \cite{ClerkRMP}.  

The case where the input signal $\hz$ is the spin operator of a qubit is also interesting.  Here, the quantum limit on QND qubit detection involves the measurement rate $\Gamma_{\rm meas}$ and the measurement-induced backaction dephasing rate $\Gamma_{\varphi}$ \cite{Devoret00,Clerk03}:
\begin{eqnarray}
	\Gamma_\varphi \geq \Gamma_{\rm meas},
	\label{eq:QubitQL}
\end{eqnarray}
where for weak coupling:
\bse
\label{eqs:QubitRates}
\begin{eqnarray}
	\Gamma_{\varphi} =  2 \bS_{FF}[0] / \hbar^2
		\label{eq:QubitDephasing}\\
	\Gamma_{\rm meas} = \left( 2 \bS_{zz}[0] \right)^{-1}.  \label{eq:QubitMeas}
\end{eqnarray}
\ese
Thus, reaching the quantum limit on QND qubit detection places more stringent requirements on the detector than those required to have a quantum-limited noise temperature:  not only must the quantum noise bound of Eq.~(\ref{eq:QNConstraintZ}) be satisfied as an equality, but in addition, there must be no backaction-imprecision correlations (e.g.~$\bS_{zF}[0] = 0$). 

In the following sections, we will calculate the nonlinear cavity's noise and response functions, and determine whether it reaches the quantum limit on its noise temperature, and on QND detection.  We note in passing that Ref. \cite{Blencowe08} also addresses the quantum limit on amplification (specifically position detection) using a nonlinear cavity in a similar regime to that considered here.  Their analysis is based on alternative formulation of the quantum limit which is not equivalent to the one discussed here; in particular, they did not address whether the nonlinear cavity optimizes the quantum noise inequality of Eq.~(\ref{eq:QNConstraintZ}), or consider its noise temperature as defined in Eq.~(\ref{eq:TNDefn}).

\section{Behaviour near bifurcation}
\label{sec:Mapping}

\subsection{Mapping to a degenerate parametric amplifier}

We begin our analysis by using standard input-output theory \cite{Gardiner85, Gardiner00} to derive the Heisenberg equation of motion for the cavity field, in the absence of any coupling to the signal:
\ba
	\frac{d}{dt} \ha = -\frac{i}{\hbar} \left[ \ha, H_{\rm sys} \right] -\frac{\kappa}{2} \ha - \sqrt{\kappa} \hbin(t).
	\label{eq:CavityEOM}
\ea
Here, $\hbin(t) = \bar{b}_{\rm in} e^{-i \omega_d t}  + \hxi(t)$  describes the input field incident on the cavity from the transmission line;  its average value $\bar{b}_{\rm in}$ describes the coherent drive applied to the cavity at frequency $\omega_d = \omega_c + \Delta$, while $\hxi(t)$ describes quantum and classical noise entering the cavity from the drive port.  Without loss of generality, we take the drive amplitude $\bar{b}_{\rm in}$ to be real and positive. 


We are interested in driving strengths that result in a large average number of quanta $\nbar$ in the cavity, but at the same time are not so strong that there is multistability in the classical cavity dynamics.  It is thus useful to write the cavity anhiliation operator $\ha$ as the sum of a classical and quantum part: this takes the form 
\be
	\ha(t) = e^{-i \omega_d t} e^{i \phi_a} \left( \cavityavg + e^{i \pi/4} \hd(t) \right).
	\label{eq:CavityDecomp}
\ee  

The complex number $e^{i \phi_a} \cavityavg \equiv \langle \ha \rangle $ is simply determined by the classical equations of motion, whereas $\hd$ describes the influence of classical and quantum noise (and eventually, the coupling to the input signal). We have chosen the phase of the second term in Eq.~(\ref{eq:CavityDecomp}) to simplify the following analysis.  From Eq.~(\ref{eq:CavityEOM}), we find that the average cavity photon number $\cavityavg$ is determined by the classical equation
\be
	\label{eq:bif_eta}
	\photonavg 
	\left[
		(\kappa/2)^2+(2\nonlinearconst \photonavg+\Delta)^2
	\right] = \kappa \left( \bar{b}_{\mathrm{in}} \right)^2.
\ee

We can now use Eq.~(\ref{eq:CavityEOM}) to write an equation for $\hd$; retaining only leading terms in $\nbar \gg 1$ yields a linear equation:
\ba
	\frac{d}{dt} \hd = -\frac{i}{\hbar} \left[ \hd, H_{\rm dpa} \right] -\frac{\kappa}{2} \hd - \sqrt{\kappa} \hxi(t),
	\label{eq:CavitydEOM}
\ea
where
\be
\label{eq:hamiltonian_dpa}
\hH_{\rm dpa}= -\hbar \tdel \:\hddag \hd +  i \hbar \frac{ \tlam }{2} \left( \hddag \hddag - \hd \hd \right).
\ee
Eq.~(\ref{eq:hamiltonian_dpa}) is simply the Hamiltonian of a degnerate parametric amplifier (DPA) driven by a non-resonant pump, where a single pump mode photon can be converted into two ``signal" mode photons and vice-versa (see, e.g.~\cite{Gardiner00}).  Here, the classical cavity field $\bar{a}$ plays the role of the pump mode, while the displaced cavity field $\hd$ plays the role of the ``signal" mode.  The effective parametric interaction strength $\tlam$ and effective pump detuning $\tdel$ are given by:
\bse
\label{eqs:DPAParams}
\ba
	\tlam & = &  2\nonlinearconst \photonavg, \\
	\tdel & = & \Delta+4\nonlinearconst \photonavg.
\ea
\ese

The above mapping of the driven nonlinear cavity to a detuned DPA is general, and only relies on $\nbar \gg 1$.  We will be especially interested in operating points near the point of bifurcation, as these allow a maximal amplifier gain.  As one approaches the bifurcation the effective DPA parameters $\tlam, \tdel$ approach universal values.  To see this, note first that a standard analysis of the classical equations of motions shows that the bifurcation occurs at a critical drive amplitude $\bar{b}_{\rm in, bif}$ satisfying:
\ba
	\label{eq:force_bifurcation}
	\left[ \bar{b}_{\rm in, bif} \right]^2 
		& = &
			\frac{1}{6\sqrt{3}} \frac{\kappa^2}{\nonlinearconst} .
\ea
For $\bar{b}_{\rm in} <  \bar{b}_{\rm in, bif}$, $\nbar$ is a single-valued function of $\Delta$.  For $\bar{b}_{\rm in} = \bar{b}_{\rm in, bif}$,
the slope of $\nbar$ versus $\Delta$ is infinite at a single point $\Delta = \Delta_{\rm bif}$; one finds from 
Eq.~(\ref{eq:bif_eta})   
\bse
\ba
	\label{eq:condition_bif}
	\Delta_{\mathrm{bif}} &=& -\frac{\sqrt{3}}{2}\kappa \\
	\label{eq:condition_bif2}
	 \photonavg_{\mathrm{bif}} &=& \frac{1}{2\sqrt{3}}\frac{\kappa}{\nonlinearconst}.
\ea
\ese

It thus follows from Eqs.~(\ref{eqs:DPAParams}) that the parameters of the effective DPA attain universal values at the bifurcation:
\bse
\label{eqs:DPABifParams}
\ba
	\tlam_{\mathrm{bif}}
	&=& \frac{\kappa}{\sqrt{3}} \\ 
	\tdel_{\mathrm{bif}} 
	&=& \frac{\kappa}{2\sqrt{3}}.
	\label{eq:DPADeltaBif}
\ea
\ese
Note crucially that for cavity operating points near the bifurcation, {\it the effective DPA pump detuning $\tdel$ is nonzero}.  As we will see in the next subsection, this will have a pronounced impact: {\it the amplified and squeezed quadratures of the DPA are not orthogonal}.  This in turn has a significant effect on the noise properties of the nonlinear cavity detector.


\begin{figure}[t!]
	\centering
	\includegraphics[width= 0.99 \columnwidth]{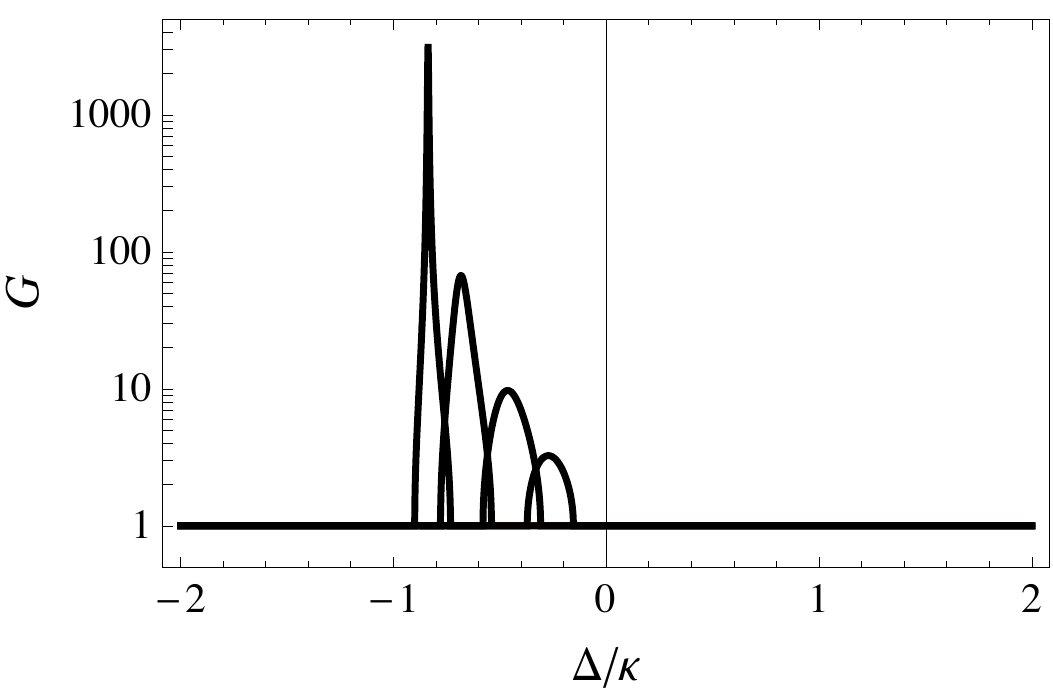}
	\caption{
		Parametric, zero-frequency photon number gain $G \equiv G[0]$ versus drive detuning $\Delta$, for various driving strengths $\bar{b}_{\rm in} < \bar{b}_{\rm in,bif}$ 
		(same values as in Fig.~\ref{fig:NBar}).  The gain is maximized near values of the detuning where the slope $d \bar{n} / d \Delta$ is maximal.}
	\label{fig:Gain}
\end{figure}

\subsection{Amplified cavity quadrature}

To appreciate the implications of pump detuning in our effective paramp model, we consider the equations of motion corresponding to Eq.~(\ref{eq:hamiltonian_dpa}).
We will be interested throughout in parameter regimes where this effective paramp has a photon number gain larger than one; 
this necessarily requires $\tlam > | \tdel |$.  In such regimes, the analysis is most conveniently presented by first introducing canonically conjugate quadrature operators $\hX$ and $\hP$:
\bse
\label{eqs:NewQuads}
\ba
\hX &=& \frac{1}{\sqrt{2}}\big(e^{-i \theta/2 }\;\hd+ e^{i \theta/2 } \;\hddag \big)  
	\label{eqs:NewQuadsX} 	\\
\hP &= &\frac{-i}{\sqrt{2}}\big(e^{-i \theta/2 }\;\hd- e^{i \theta/2 } \;\hddag \big)
\label{eqs:NewQuadsP}
\ea  
\ese 
where for the angle $\theta$ ($-\pi/2 \leq \theta \leq \pi/2$) is given by:
\be
\label{eq:theta_def}
	\sin{\theta} = \tdel / \tlam.
\ee
As we will see, the above definition ensures that $\hX$ is the quadrature amplified by the cavity.   We also define corresponding quadratures $\hX_{\rm in}$ and $\hP_{\rm in}$ of the operator $\hxi$ associated with noise entering the drive port (e.g. these are defined by substituting $\hd \ra \hxi$ in Eqs.(\ref{eqs:NewQuads}))

With these definitions, the equations of motion are easily solved upon Fourier transforming (see Appendix \ref{sec:appendix_dpa}):
\bse
\label{eqs:EOMSols}
\begin{eqnarray}
	\label{eq:eqns_noz}
	\hX[\omega] &=& 
			-\skap \Big( 
				\,\chi_1[\omega]\,\hX_{\mathrm{in}}[\omega] \nonumber 
					\label{eq:XEOM}\\
	&&
			- 	 \tan{\theta} \left( \chi_1[\omega] - \chi_2[\omega] \right) \, \hP_{\mathrm{in}}[\omega]
		\Big),   \\
	\hP[\omega] &=& 
		-\skap\, \chi_2[\omega]\, \hP_{\mathrm{in}}[\omega],	\label{eq:PEOM}
\end{eqnarray}
\ese
where the susceptibilities $\chi_1, \chi_2$ are given by:
\bse
	\label{eqs:Chis}
\ba
	\chi_1[\omega]&=& 
		\left(- i \omega + \kappa/2-\sqrt{\tlam^2-\tdel^2}  \right)^{-1}  \\
	\chi_2[\omega]&=& 
		\left( -i \omega + \kappa/2+\sqrt{\tlam^2-\tdel^2}  \right)^{-1}. 
\ea
\ese
%
%

For the case of a resonant pump (i.e.~$\tdel = \theta = 0$), these equations take a simple form and describe the usual behaviour of a DPA:  as 
$\tlam$ approaches $\kappa/2$ from below (the parametric threshold), $\kappa \chi_1[0] \rightarrow \infty$, $\kappa \chi_2[0] 
\rightarrow 1$, and  $\hX$ ($\hP$) is the amplified (squeezed) quadrature.  By considering quadratures of the output field leaving the cavity, one finds that the photon number gain for the $X$ quadrature is given by:
\be
\label{eq:gain_def}
	G [\omega] \equiv \left| 1- \kappa \chi_1[\omega] \right|^2
.\ee
We will refer to $G[\omega]$ as the ``parametric gain" of our system in what follows.
For a resonant pump, the amplified and squeezed quadratures are clearly orthogonal (i.e.~canonically conjugate).  
Note that $G[\omega]$ has a Lorentzian form, implying that there is only appreciable gain for frequency in a bandwidth $\Bdpa$, 
\begin{eqnarray}
	\Bdpa \equiv 1/ \chi_1[0].
	\label{eq:DPABandwidth}
\end{eqnarray}
We refer to $\Bdpa$ as the ``parametric bandwidth" in what follows; in the large parametric gain limit, $\Bdpa \sim \kappa / \sqrt{G[0]}$.

The situation is more involved in the case of interest here, where the effective pump is not resonant
(c.f. Eqs.~(\ref{eq:DPADeltaBif})), and hence $\theta \neq 0$.  We still have a parametric threshold when $\tlam$ approaches 
$\sqrt{  \kappa^2/4 + \tdel^2}$ from below; as before, $\kappa \chi_1[0] \rightarrow \infty$ in this limit while $\kappa \chi_2[0] \rightarrow 1$.  It is easy to verify from Eqs.~(\ref{eqs:DPABifParams})
 that the parametric threshold coincides with the cavity bifurcation.  
 It also follows from Eqs.~(\ref{eqs:EOMSols}) that for any pump detuning $\tdel$, $\hX$ is the amplified quadrature:  
noise (or signal) incident in the $X$ quadrature (i.e.~$\hX_{\rm in}$) {\it only} drives the $X$ cavity quadrature, and is multiplied by the large susceptibility $\chi_1$.  The photon number gain for signals in the $X$ quadrature 
continues to be described by Eq.~(\ref{eq:gain_def}); as expected, this gain diverges as one approaches the bifurcation (see Fig.~\ref{fig:Gain}).  As one approaches the bifurcation, Eqs.~(\ref{eqs:DPABifParams}) imply that the angle $\theta$ which defines $\hX$ takes the universal value:
\begin{eqnarray}
	\theta_{\rm bif} = \pi/6.
\end{eqnarray}

More troublesome when $\tdel \neq 0$ are the dynamics of the $P$ quadrature, the quadrature orthogonal to the amplified quadrature.  For a non-zero detuning, $P$ is {\it not} the squeezed quadrature.  Noise or signals incident on the cavity in the $P$ quadrature (i.e. $\hP_{\rm in}$) appear both in the cavity $P$ quadrature (where it is multiplied by the small susceptibility $\chi_2$), {\it as well} as in the cavity $X$ quadrature, where it is also ``amplified" (i.e.~multiplied by the large susceptibility $\chi_1$).

To summarize, we have shown that near the bifurcation, the driven nonlinear cavity of Eq.~(\ref{eq:ham_cavity}) maps onto a DPA with a non-zero pump detuning $\tdel$.  This non-zero detuning means that the dynamics does not correspond to the simple situation of canonically-conjugate amplified and squeezed quadratures.  As we will see, this lack of orthogonality will have pronounced implications on the cavity noise properties near the bifurcation.

\subsection{Coupling to signal and cavity gain}

To complete our mapping of the nonlinear cavity detector to a DPA, we need to restore the signal-detector coupling Hamiltonian and consider the forward gain $\chi_{IF}(t)$ of the system (c.f.~Eq.(\ref{eq:ChiIFDefn})).  This forward gain tells us how strongly the input signal $\hz$ influences the output homodyne current, and will {\it not} be identical to the parametric photon number gain $G$ discussed above.  While one could calculate $\chi_{IF}[\omega]$  directly using a Kubo formula, it is simpler here to simply re-derive the equations of motion for the cavity field including the coupling to $\hz$.  Retaining only leading-order terms in $\nbar$, the signal-cavity coupling Hamiltonian $H_{\rm int}$ in Eq.~(\ref{eq:Hint}) retains the form $\hH_{\rm int} = \hF \cdot \hz$, with the generalized force operator $\hF$ taking the form:
\be
	\label{eq:force}
	\hat{F} \equiv  A \ha^\dagger \ha \simeq	
	 ( \sqrt{2 \nbar} A ) (\sin{\nu} \hat{X} +\cos{\nu} \hat{P}),
\ee
where
\be
\label{eq:nu}
\nu \equiv \theta/2 + 3 \pi/4.
\ee
We have dropped a constant term in $\hF$ which can absorbed into the Hamiltonian of the signal source.  
We see that in general, the input signal $\hz$ couples to both $\hX$ and $\hP$, and thus will enter the linearized cavity equations of motion as a driving term for both these quadratures.  To be explicit, one should make the following replacements in Eqs.(\ref{eqs:EOMSols}):
\bse
\label{eqs:SignalInputs}
\begin{eqnarray}
	\hX_{\rm in}[\omega] & \rightarrow &
		 \hX_{\rm in}[\omega] - \left(\sqrt{\frac{2 \nbar}{\kappa}} A \cos \nu \right)  \hz[\omega] \\
	 \hP_{\rm in}[\omega] & \rightarrow &
		 \hP_{\rm in}[\omega] + \left( \sqrt{\frac{2 \nbar}{\kappa}} A \sin \nu \right)  \hz[\omega].
\end{eqnarray}
\ese
Note that at the bifurcation, the angle $\nu$ takes on the universal value:
\begin{eqnarray}
	\nu_{\rm bif} = \pi/ 12 + 3 \pi/4 = 5 \pi/6 =\pi  - \theta_{\rm bif}.
\end{eqnarray}

Having characterized the signal-detector coupling, we now turn to the output homodyne current (c.f.~Eq.~(\ref{eq:IDef})).  This current is essentially one quadrature of the cavity output field, and may be written:
\begin{eqnarray}
	\hI[\omega] & = & \sqrt{\kappa}\left( \cos \varphi_h  \hX_{\rm out}[\omega] +  \sin \varphi_h  
	\hP_{\rm out}[\omega] \right),
	\label{eq:IQuads}
\end{eqnarray}
where $\varphi_h$ is determined by the phase of the reference beam used in the homodyne measurement, and the output operators are given by the standard input-output relations  \cite{Gardiner85, Gardiner00}, e.g.~:
\begin{eqnarray}
	\hX_{\rm out}[\omega] = \hX_{\rm in}[\omega] + \sqrt{\kappa} \hX[\omega]
\end{eqnarray}
It thus follows from Eqs.~(\ref{eqs:EOMSols}) and (\ref{eqs:SignalInputs}) that the linear-response gain of the cavity amplifier will have the general form:
\begin{eqnarray}
	\chi_{IF}[\omega]	& = & 
		 \hbar A \sqrt{2 \nbar} \kappa  \left( \lambda_1 \cdot \chi_1[\omega]  + 
			\lambda_2 \cdot \chi_2[\omega]  \right),
	\label{eq:ChiIF}
\end{eqnarray}
where 
\bse
\begin{eqnarray}
	\lambda_1 & = &
		\cos \varphi_h \left( \cos \nu + \tan \theta \sin \nu \right) \\
	\lambda_2 & = &
		-\sin \nu \left( \cos \varphi_h \tan \theta + \sin \varphi_h \right).
\end{eqnarray}		
\ese

\section{Amplifier noise in the large gain, low frequency limit}
\label{sec:LowFrequencyAnalysis}

We are most interested in the properties of our cavity amplifier close to the bifurcation, where the parametric gain defined in Eq.(\ref{eq:gain_def}) satisfies $G[0] \equiv G \gg 1$.  In this regime, 
we expect amplification of input signals in a narrow band of frequencies $\omega < \Bdpa \sim \kappa / \sqrt{G}$.  We thus begin our analysis by considering the amplifier noise to leading order in the large parameter $G$ and for
$\omega \ll \Bdpa$; the latter condition allows us to take the zero-frequency limit of cavity noise and response functions.  
We also assume the ideal case where the cavity is only driven by vacuum noise.  Note that 
it is straightforward to use Eq.~(\ref{eq:bif_eta}) to determine how $G$ behaves as a function of driving strength as one approaches the bifurcation from below.  Assuming that the drive detuning $\Delta$ is always chosen in order to maximize $G$, one finds that near the bifurcation:
\begin{equation}
	G \sim 4 \left(  \frac{  |\bar{b}_{\rm in,bif}|^2   }{    | \bar{b}_{\rm in,bif} |^2 - |\bar{b}_{\rm in}|^2  } \right)^3.
\end{equation}

We start with the amplifier's forward gain.  In the limit $G \rightarrow \infty$.  Eq.~(\ref{eq:ChiIF}) yields:
\begin{eqnarray}
	\chi_{IF}[0]	& \sim & 
		\sqrt{2 \nbar \kappa } A \lambda_1\cdot \sqrt{G}.
		\label{eq:ChiIFLeading}
\end{eqnarray}
As expected, the forward gain is (to leading order) proportional to square root of the DPA photon number gain.

Turning to the cavity output noise, we note that for $G \gg 1$ and for small frequencies, Eqs.~(\ref{eqs:EOMSols}) yields that the cavity $\hP$ quadrature is negligible in comparison to the $\hX$ quadrature.  As such, we can drop the second term in Eq.~(\ref{eq:IQuads}), and treat the homodyne current operator $\hI$ as being proportional to $\hX_{\rm out} \propto \hX \propto \sqrt{G}$, Thus, for $G \rightarrow \infty$:
\begin{eqnarray}
	\hI[\omega]  
			& \sim & \kappa \cos \varphi_h \hX[\omega].
			\label{eq:ILeading}
\end{eqnarray}

Looking at Eq.~(\ref{eq:force}) for the backaction force operator $\hF$, we see a similar argument holds.  Thus,  to leading order in $G$, $\hF$ is also proportional to $\hX \propto \sqrt{G}$:  
\begin{eqnarray}
	\hF[\omega] \sim \sqrt{2 \nbar} A \sin \nu  \hX[\omega].
	\label{eq:FLeading}
\end{eqnarray}
Thus, to leading order in $G$, {\it backaction and imprecision noises are perfectly correlated with one another}, as they only differ by a constant.  Their spectral densities will simply be proportional to the spectral density of the 
amplified cavity quadrature, $\hX$.

\subsection{Imprecision Noise}

\begin{figure}[t!]
	\centering
	\includegraphics[width= 0.85 \columnwidth]{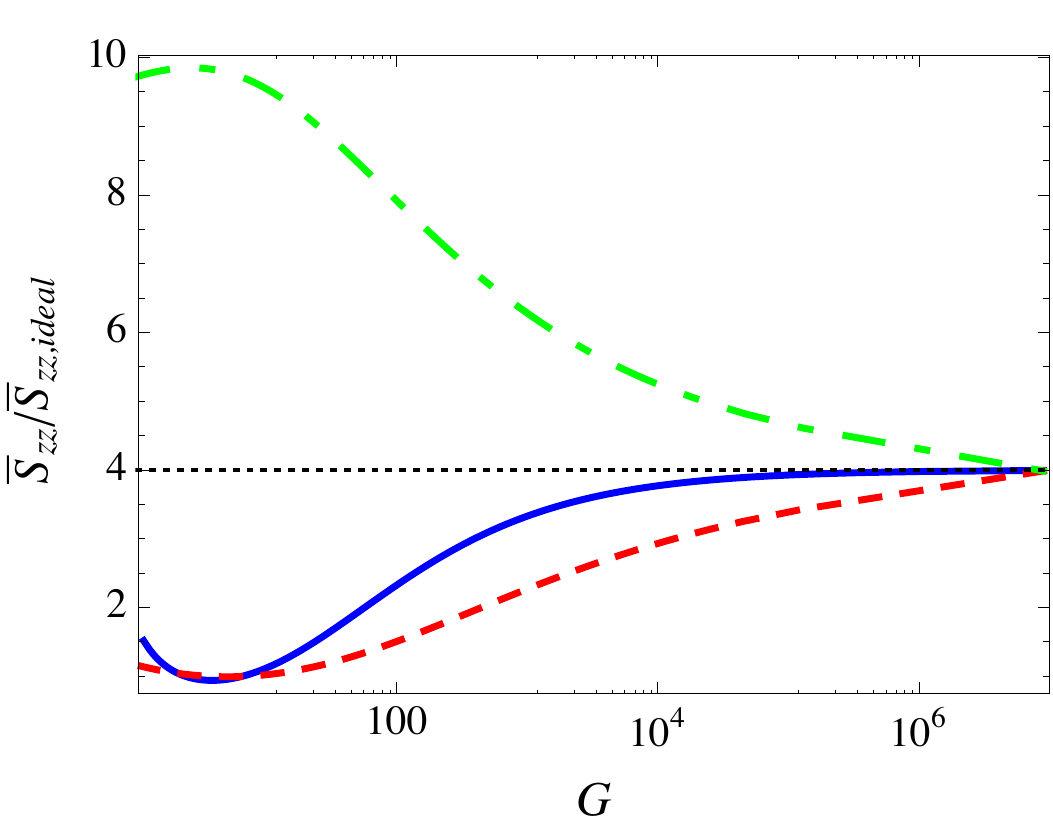}
	\caption{
		Imprecision noise $S_{zz}[0]$, scaled by the imprecision noise of an ideal degenerate parametric amplifier, $\S_{zz,{\rm ideal}}$ versus 
		parametric photon number gain $G$.  For large $G$, the imprecision noise is four times the ideal value.  
		 Solid blue:  increase $G$ by increasing the drive amplitude $\bar{b}_{\rm in}$ towards $\bar{b}_{\rm in, bif}$, 
		using an optimal $\Delta$ for each drive strength.  Dashed red:  fix $(\bar{b}_{\rm in} /  \bar{b}_{\rm in, bif} )^2= 0.995  $, increase $G$ by tuning $\Delta$ to approach
		the optimal value from below.  Long dash - short dash, green:  same as previous, but increase $G$ by tuning $\Delta$ to approach
		the optimal value from above.}
	\label{fig:Szzplot}
\end{figure}

The leading-order-in-$G$ intrinsic output noise of the amplifier (i.e.~noise in the homodyne current at $A=0$) thus follows easily from Eq.~(\ref{eq:ILeading}) and (\ref{eq:XEOM}) (see Appendix \ref{sec:appendix_dpa}).  In the low frequency limit, we have:
\ba
	\bS_{II}[0]  
	& \sim & 
		\frac{\kappa}{2} \cos^2 \varphi_h \cdot G
		  (1 + \tan^2{\theta}). 
		\label{eq:sii}
\ea
The two terms in the last factor represent two distinct physical contributions to the output noise.  The $\theta$-independent term arises from $X$-quadrature input noise ($\hX_{\rm in}$) being amplified and appearing in $\hX$.  In contrast, the term proportional to $\tan^2 \theta$ is a direct consequence of the non-zero effective pump detuning $\tdel$.  The resulting non-orthogonality of amplified and squeezed quadratures causes $P$-quadrature input noise ($\hP_{\rm in}$) to {\it also} be amplified and appear in the amplifier output $\hI$.  We thus see that $\tdel \neq 0$ causes the output noise to be larger than what would be expected for a resonantly-pump DPA with equivalent photon number gain $G$.

Combing the above expression with Eq.~(\ref{eq:ChiIFLeading}) for the gain, we find that the low-frequency imprecision noise 
(c.f.~Eq.(\ref{eq:ImprecisionNoiseDefn}) )near the bifurcation (i.e.~$G \rightarrow \infty$) is given by:
\ba
	\label{eq:imprecision_noise}
	\bS_{zz}[0]  
	& \sim & \bS_{zz,{\rm ideal}}[0] \cdot h_{\rm imp}(\theta, \nu),
\ea
where
\bse
\label{eqs:IdealDPA}
\ba
	\bS_{zz,{\rm ideal}} & = & 		
		\left(  \frac{ \hbar^2 \kappa}{4 \photonavg A^2  }  \right) \\
	h_{\rm imp}(\theta,\nu) & = &
		 \frac{ 1+\tan^2{\theta}  }{(\cos{\nu} +\sin{\nu}\tan{\theta} )^2}. 
\ea		
\ese
Here, $\bS_{zz,{\rm ideal}}$ is the imprecision noise of an ``ideal" DPA in the large gain limit.  By ``ideal", we mean a DPA which was pumped on resonance, and where the input signal $z$ only drives the amplified $X$ quadrature (i.e.~the angle $\nu$ in Eq.(\ref{eq:force}) would be zero).  $h_{\rm imp}(\theta,\nu) > 1$ describes the increase of $\bS_{zz}$ due to the fact that our nonlinear cavity does not realize a DPA in this ideal fashion.  The numerator of $h_{\rm imp}$ describes the extra output noise due to the non-resonant effective pump, as discussed after Eq.~(\ref{eq:sii}).  The denominator describes the reduction in gain coming from the fact that the signal  drives both the $\hX$ and $\hP$ quadratures.

Finally, we can further simplify our result by using the fact that near the point of bifurcation, the effective DPA parameters approach universal values (c.f.~Eq.~(\ref{eqs:DPABifParams})).  To leading order, we can simply replace $\theta$ by 
its value at the bifurcation $\theta_{\rm bif} = \pi/6$.  We thus obtain our final expression for the imprecision near the bifurcation:
\be
\label{eq:szz_ideal}
	\bS_{zz}[0] \sim 
		4 \bS_{zz, \mathrm{ideal}}.
\ee
{\it In large gain limit, the imprecision of the nonlinear cavity amplifier is a factor of four times what would be expected from a theoretically ideal degenerate parametric amplifier.}  The behaviour of the imprecision noise relative to the ideal value is shown in Fig.~\ref{fig:Szzplot}.

\begin{figure}
	\centering
	\includegraphics[width= 0.85 \columnwidth]{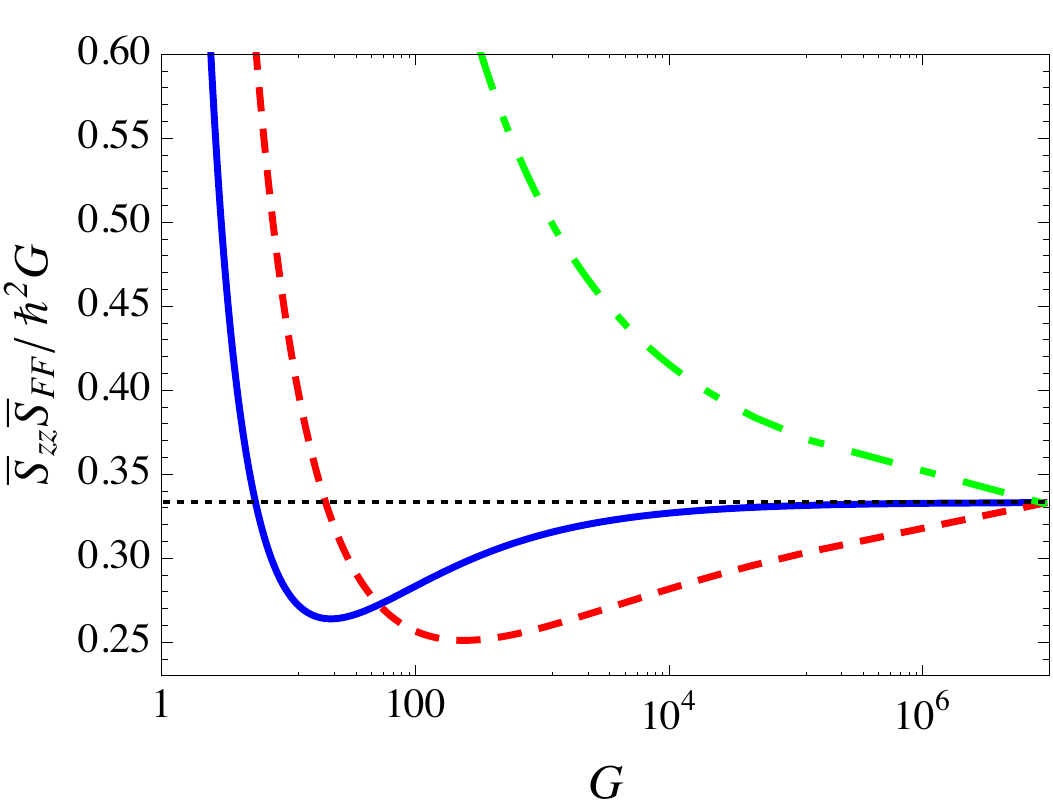}
	\caption{
		Backaction-imprecision product $\bS_{zz}[0] \bS_{FF}[0]$ (scaled by $\hbar^2 G$) versus parametric photon number gain $G$, demonstrating
		the universal scaling predicted for large $G$.  
		 Individual curves correspond to the same parameters
		as in Fig.~\ref{fig:Szzplot}.  }
	\label{fig:BAImpUniversal}
\end{figure}

\begin{figure} 
	\centering
	\includegraphics[width= 0.95 \columnwidth]{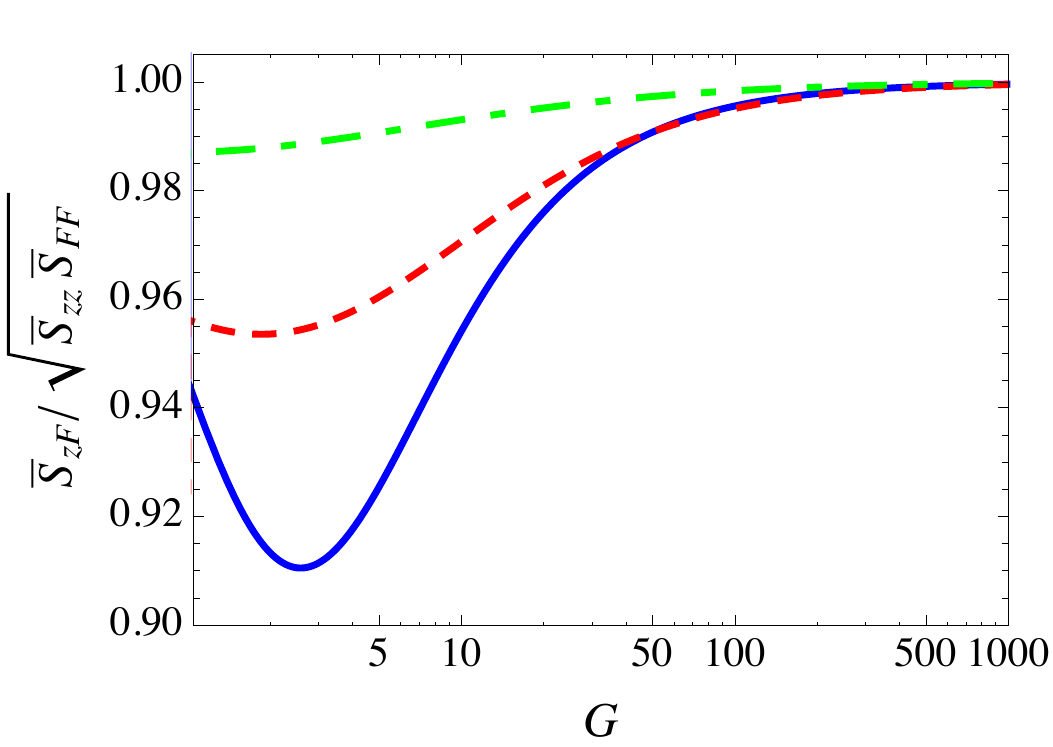}
	\caption{
		Backaction-imprecision correlations, as measured by $\bS_{zzF}[0] / \sqrt{ \bS_{zz}[0] \bS_{FF}[0] }$ versus parametric photon number gain $G$.
		As expected, the curves all tend to the universal value of $1$ as $G \ra \infty$.  Individual curves correspond to the same parameters
		as in Fig.~\ref{fig:Szzplot}.}
	\label{fig:SzFUniversal}
\end{figure}

\subsection{Backaction noise and backaction-imprecision product}

From Eq.~(\ref{eq:FLeading}), we see that to leading-order-in-$G$, the low-frequency backaction noise spectral density 
$\bS_{FF}[0]$ will just be proportional to the low-frequency output noise spectral density $\bS_{II}[0]$.  Using the universality of the DPA parameters
near the bifurcation, we find that in the $G \ra \infty$ limit:
\begin{eqnarray}
	\label{eq:backaction}
	S_{FF}[0] & \sim & 
		 \frac{1}{3} \frac{A^2 \nbar}{\kappa} G.
		\label{eq:SFFLeading}
\end{eqnarray}
We see the backaction diverges as the parametric photon number gain $G$; this is a simple consequence of the fact that our dispersive coupling unavoidably leads the signal $\hz$ to be coupled to the amplified cavity quadrature $X$.  The full expression (valid for arbitrary $G$) is not too unwieldy and is given in the Appendix as Eq.~(\ref{eq:specFF}).  Note that similar spectral densities for a nonlinear cavity were calculated using a linearized Fokker-Plank approach in Ref.~\cite{Wilhelm10} and (in the classical, high temperature regime) Ref.~\cite{Dykman94}.

Combining our results, we see that near the bifurcation, the backaction - imprecision product will be much larger than the minimum value of $\hbar^2/4$ allowed by quantum mechanics.  In the large $G$ limit, we have:
\begin{eqnarray}
	\bS_{FF}[0] \bS_{zz}[0] \sim \frac{G}{3} \hbar^2. 
	\label{eq:BAImprecisionLeading}
\end{eqnarray}
Figure \ref{fig:BAImpUniversal} shows the scaling of $\bS_{FF}[0] \bS_{zz}[0]$ versus parametric gain $G$ as one approaches the bifurcation by either tuning the drive detuning $\Delta$ or the drive strength $\bar{b}_{\rm in}$; the universal asymptotic behaviour described by Eq.~(\ref{eq:BAImprecisionLeading}) is clear.

The above result implies that if one cannot make use of backaction-imprecision noise correlations, one is very far from having a quantum limited device.  In particular, {\it near the bifurcation the nonlinear cavity detector cannot be used for QND qubit detection}:  in such an experiment, the backaction dephasing rate will be a factor $G \gg 1$ larger than the minimum rate dictated by quantum mechanics (c.f.~Eq.~(\ref{eq:QubitQL})).  Note that the situation is very different for a linear cavity:  there, as long as one drives the cavity on resonance, the $\bS_{FF}[0] \bS_{zz}[0]$ product attains the minimum possible value of $\hbar^2 / 4$ \cite{ClerkRMP, Blais04}.

It is tempting to think that by simply changing the cavity operating point slightly, one could achieve a situation where the input signal is only coupled to the cavity quadrature $\hP$, and thus avoid the problematic diverging backaction found above.  From Eqs.~(\ref{eq:force}) and (\ref{eq:nu}), we see that this would require an operating point for which the angle $\theta = \pi/2$.  However, from Eqs.~(\ref{eq:theta_def}) and (\ref{eqs:Chis}), this in turn implies that the cavity would have no parametric gain:  $G=1$.  Thus, one cannot solve the problem of large backaction by simply changing the drive detuning without simultaneously getting rid of the amplifier gain.

\subsection{Comparison with linear-cavity backaction formula}

For a linear cavity, one can directly connect the backaction noise spectral density at zero frequency to how strongly the average cavity amplitude $\langle \ha \rangle$ changes in response to a change in the signal.  One finds:
\begin{eqnarray}
	\bS_{FF,{\rm lin}}[0] & = & A^2 \left|  \frac{d \langle \ha \rangle  }{d \Delta} \right|^2.
	\label{eq:SFFOverlap}
\end{eqnarray}
This elegant result was first derived in Ref.~\cite{Gambetta06} in the case where $\hz$ is a spin operator for a qubit; $\bS_{FF}[0]$ in this case is directly proportional to the qubit dephasing rate (c.f.~Eq.(\ref{eq:QubitDephasing})).  Heuristically, it expresses the fact that the backaction disturbance of the measurement is directly related to the distinguishability of cavity states associated with different values of the input signal.  A small change in the input signal causes a small displacement of the coherent state describing the cavity.   Eq.~(\ref{eq:SFFOverlap}) implies that the backaction dephasing $\Gamma_{\varphi}$ (and hence $\bS_{FF}$) is directly determined by the overlap between this displaced coherent state and the original coherent state describing the cavity.

One would not expect  Eq.~(\ref{eq:SFFOverlap}) to apply in general to our nonlinear cavity detector, as now the intracavity state corresponding to a given fixed value of the input signal is {\it not} a coherent state, or even a pure state \cite{Milburn81}.  This is a direct result of the squeezing and amplification of the cavity noise that occurs as one approaches the bifurcation.  However, if one is far from the bifurcation, these effects should be minimal, and one might expect Eq.~(\ref{eq:SFFOverlap}) to remain valid.  This idea was recently put forward in  Ref.~\cite{Esteve10}, and derived within an approximation which neglects noise squeezing of the cavity.  Our approach fully accounts for the squeezing of the intracavity fluctuations, and allows us to test the general validity of Eq.~(\ref{eq:SFFOverlap}).  Surprisingly, we find that this expression {\it exactly} captures the full backaction noise, even close to the bifurcation:
\begin{eqnarray}  
	\bS_{FF,{\rm lin}}[0] & = \bS_{FF}[0].
	\label{eq:SFFApprox}
\end{eqnarray}
Here $\bS_{FF}[0]$ is the full expression for the backaction noise spectral density that follows from Eqs.~(\ref{eqs:EOMSols}), (see  Eq.~(\ref{eq:specFF})).  We see that despite the fact that the cavity is not in a coherent state or even a pure state,  Eq.~(\ref{eq:SFFOverlap}) remains valid for the nonlinear cavity amplifier; that this should be so is by no means  {\it a priori} obvious.

Ref.~\cite{Esteve10} also suggests that the nonlinear cavity detector reaches the quantum limit on QND detection (c.f.~Eq.~(\ref{eq:QubitQL})), implying that the backaction-imprecision product $\bS_{FF} \bS_{zz}$ attains its minimum possible value of $\hbar^2 / 4$.  In contrast, we find that the backaction noise $\bS_{FF}$ (and hence backaction dephasing rate) is factor $G \gg 1$ larger than the quantum limited value (c.f.~Eq.(\ref{eq:BAImprecisionLeading}) and Fig.~\ref{fig:BAImpUniversal}).  The discrepancy here arises from the fact that Ref.~\cite{Esteve10} does not explicitly calculate the measurement rate (i.e.~$1/\bS_{zz}$) for a specific, optimized cavity readout scheme, but rather assumes that it also be given (up to a prefactor) by overlap expression in Eq.~(\ref{eq:SFFOverlap}).  This would imply the measurement imprecision noise $\bS_{zz}[0]$ scales like $1/G$ in the large-$G$ limit.  In contrast, we explicitly consider homodyne detection of the cavity output.  We find that the imprecision noise (and hence measurement rate) are independent of $G$ in the large gain limit, in agreement with Ref.~\cite{Vijay10}.  This is a simple consequence of the fact that the nonlinear cavity's parametric gain amplifies both the signal {\it and} the vacuum fluctuations driving it by the same factor of $\sqrt{G}$.

\subsection{Quantum limit on the amplifier added noise}

While in the low-frequency, large gain limit, the nonlinear cavity system cannot function as a quantum-limited QND qubit detector, it may nonetheless be a quantum limited linear amplifier (i.e.~have the minimum noise temperature $T_N$ allowed by quantum mechanics).  This difference stems from the fact that when used as an amplifier (in the ``op-amp" mode), one can take advantage of correlations between backaction and imprecision noise by tuning the susceptibility of the signal source (e.g.~in a voltage amplifier, the source impedance).  

To leading order in $G$ and at low frequencies, we have shown that the backaction and output noise operators are proportional to one another, implying perfect correlation:
\be
	\bS_{zF}[0] \sim  \sqrt{ \bS_{zz}[0] \bS_{FF}[0] }.
	\label{eq:PerfectCorr}
\ee
Fig.~\ref{fig:SzFUniversal} shows the behaviour of these correlations versus $G$, where $G$ is tuned is various ways; the asymptotic, perfect correlation behaviour is clear.

Turning to Eq.(\ref{eq:TNDefn}) for the optimized noise temperature $T_N$, we see that perfectly correlated backaction and imprecision noises do not contribute.  This implies that our leading-order-in-$G$ analysis is insufficient to determine whether $T_N$ is quantum limited:  this analysis only tells us that there is no order-$\sqrt{G}$ term in $T_N$.  To determine whether the quantum limit is reached near the bifurcation, one must go beyond leading order expressions, even though we are interested in the low frequency limit.  Such an analysis is straightforward though tedious; details are presented in Appendix \ref{sec:appendix_dpa}.  Obtaining the cavity noise correlators and forward gain exactly from Eqs.~(\ref{eqs:EOMSols}) with no large-$G$ assumption, we find that at zero frequency, {\it the nonlinear cavity detector always optimizes the quantum noise inequality of Eq.~(\ref{eq:QNConstraintZ}) (i.e.~it is satisfied as an equality)}.  As such, the minimal low-frequency noise temperature given by Eq.~(\ref{eq:TNDefn}) is indeed the quantum limited value of $\hbar \omega / 2$.  We stress that this result is completely independent of the choice of homodyne phase $\varphi_h$.  

\subsection{Utility of backaction - imprecision correlations}

As always, achieving a quantum-limited noise temperature is not simply a question of having an amplifier which saturates the fundamental quantum noise inequality of Eq.~(\ref{eq:QNConstraintZ})-- one also needs to optimally tune the susceptibility $\chi_{zz}[\omega]$ of the signal source (i.e.~the source impedance).  This optimization results in two conditions, c.f.~Eqs.~(\ref{eqs:OptimalChi}).  The magnitude condition (c.f.~Eq.~(\ref{eq:OptimalChiMag})) can always be achieved by an appropriate tuning of the signal - detector coupling $A$; it corresponds to properly balancing the relative contributions of backaction and imprecision noise to the total added noise.  In contrast, the phase condition (c.f.~Eq.~(\ref{eq:OptimalChiPhase})) cannot be achieved by simply tuning $A$.  It corresponds to optimizing $\chi_{zz}[\omega]$ to optimally make use of in-phase  backaction - imprecision correlations described by $\textrm{Re } \bS_{zF}$.  

Consider the nonlinear cavity detector in the low-frequency, large gain regime considered above.  We found that it has a maximal value of correlations $\bS_{zF}$, Eq.~(\ref{eq:PerfectCorr}).  Eq.~(\ref{eq:OptimalChiPhase}) then implies that reaching the quantum limit on the noise temperature requires $\textrm{Im } \chi_{zz}[\omega]=0$.  This is in sharp contrast to the more common situation where $\bS_{zF}$ vanishes, and the optimal source susceptibility $\chi_{zz}$ must be purely imaginary.  

This has interesting consequences.  For concreteness, consider the case where our input system is a mechanical oscillator and $z$ represents a position, $\chi_{zz}[\omega]$ is simply given by:
\begin{eqnarray}
	\chi_{zz}[\omega] & = & \frac{-1/m}{\omega^2 - \omega_M^2 + i \omega \gamma}.
\end{eqnarray}
Here, $\omega_M$ is the resonance frequency of the mechanical oscillator, $m$ is its mass, and $\gamma$ is its damping rate.  We see that $\chi_{zz}[\omega]$ is purely real if one is far from resonance, i.e.~$| \omega - \omega_M | \gg \gamma$.  Thus, the nonlinear cavity detector is ideally suited to applications where one is interested in non-resonant position detection.  For example, standard interferometric  gravitational wave detectors require sensitive position detection of a test mass in the free-mass limit, i.e.~$\omega_M \ra 0$ \cite{Corbitt04}.  In this case, as long as $\omega \gg \gamma$, one always has a non-resonant situation, and $\chi_{zz}$ is real.  For such frequencies, the nonlinear cavity amplifier would be able to achieve a quantum-limited noise temperature.  In contrast, if one used a detector with $\bS_{zF}=0$ in this regime,  the noise temperature is at best a factor $\omega_M / \gamma \gg 1$ larger than the quantum limited value.  The utility of using correlations between backaction and imprecision noise is well-known in the gravitational wave community \cite{Chen01}, though it is not usually discussed in terms of the general noise temperature language used here.  

Finally, we note that if the input signal $\hz$ was a voltage, and we think of our cavity amplifier as a voltage amplifier, the 
requirement that the input susceptibility be purely real to optimize the noise temperature translates into requiring a signal source with a purely imaginary source impedance \cite{ClerkRMP}. 

\section{Amplifier noise at non-zero frequencies}
\label{sec:FiniteFrequencyAnalysis}

It is straightforward to extend our analysis to describe the amplification of signals with frequencies $\omega$ that are non-zero, but still small enough that the parametric gain $G[\omega] \gg 1$.  It follows from Eq.~(\ref{eq:gain_def}) that in the $G[0] \gg 1$ limit, this requires $\omega \leq \Bdpa \sim \kappa / \sqrt{G[0]}$.  Simple analytic expressions are easily obtained in the limit where $G[0] \rightarrow \infty$ while $ \omega /  \Bdpa$ stays finite.  To leading order in $G[0]$, one finds (as expected) that the photon number gain $G[\omega]$ and forward gain $\chi_{IF}[\omega]$ have a Lorentzian frequency dependence on a scale set by $\Bdpa$.  Letting $\tomega = \omega / \Bdpa$, we have:
\bse
\begin{eqnarray}
	G[\omega] & = &  \frac{G[0]}{1+\tomega^2} \\
	\chi_{IF}[\omega] & = &  \frac{\chi_{IF}[0] }{1- i \tomega}.
\end{eqnarray}
\ese
In the same limit, we find that the imprecision noise is frequency independent, whereas the remaining correlators also decay with frequency on a scale set by $\Bdpa$:
\bse
\begin{eqnarray}
	\bS_{zz}[\omega]  & = & \bS_{zz}[0] \\
	\bS_{FF}[\omega]  & = &  \frac{\bS_{FF}[0]}{1 + \tomega^2} \\
	\bS_{zF}[\omega]  & = & \frac{\bS_{zF}[0] }{1+i \tomega}.
\end{eqnarray}
\ese
 It immediately follows that for finite frequencies, the noise temperature behaves as:
\begin{eqnarray}
	k_B T_N[\omega] & \simeq & \hbar \omega \Bigg( 
		\sqrt{ \frac{1}{4}  + \frac{G}{3} \left(  \frac{\tomega}{ 1 + \tomega^2  }  \right)^2  }
	+	
	\label{eq:TNFiniteFreq}
	\sqrt{ \frac{G}{3} } \frac{\tomega}{1 + \tomega^2}
		\Bigg). \nonumber \\
\end{eqnarray}	
We thus see that at finite frequencies, the reduced noise temperature $2 k_B T_N / \hbar \omega$ rapidly increases as a function of frequency from the quantum-limited value of $1$;  in particular, it is already much greater than one for frequencies small enough to not appreciably reduce the gain.  
The leading correction at finite $\omega$ comes from the imaginary part of the noise cross-correlator $\bS_{zF}[\omega]$.  As discussed extensively in \cite{ClerkRMP}, such out-of-phase backaction-imprecision correlations cannot be taken advantage of by simply tuning the susceptibility of the source; as a result, their existence represents unused information, and thus leads to a departure from the quantum limit.  In principle, such correlations can be 
utilized via feedback techniques.

%


\section{Back-Action Cooling}
\label{sec:Cooling}

We have seen in the preceding analysis that near the bifurcation point, the backaction noise of the nonlinear cavity amplifier diverges; this prevents  quantum-limited amplification unless one can make use of noise correlations.  In this section, we change focus somewhat and consider the specific case where the input signal $\hz$ is the position of a mechanical resonator.  In this case, the large backaction of the nonlinear cavity may actually be useful:  it has the potential to strongly cool the mechanical resonator towards its quantum ground state. 

The topic of backaction cooling has received considerable attention in the optomechanics and electromechanics communities \cite{Marquardt09}.  It has been shown that the backaction of a {\it linear} cavity dispersively coupled to a mechanical resonator can be used to ground-state cool the mechanical resonator if one is in the so-called good cavity limit, where the mechanical frequency $\omega_M$ is much larger than the cavity damping rate $\kappa$ \cite{WilsonRae07, Marquardt07}.  This regime has been exploited in recent experiments with linear microwave cavities \cite{Teufel08, Schwab10} and optical cavities \cite{Aspelmeyer09, Wang09, Kippenberg09}.

 In the opposite regime of a low-frequency mechanical resonator ($\omega_M \ll \kappa$), cooling using a linear cavity is still possible, but the lowest achievable temperature is on the order of $T_{\rm BA} \sim \kappa / k_B$.  A crucial parameter is the backaction damping (or optical damping) rate $\gamma_{\rm BA}$ :  this is the enhanced damping of the mechanical resonator resulting from a net energy loss to the driven cavity.  The cooling power of the cavity backaction will be directly proportional to $\gamma_{\rm BA}$.  In the low frequency regime, a simple classical linear-response argument yields that for a mechanical resonator dispersively coupled to a cavity, 
 $\gamma_{\rm BA} \propto d \nbar  / d \Delta$ \cite{2004_12_HoehbergerKarrai_CoolingMicroleverNature, Marquardt07}.  
 As discussed in Ref.~\cite{Blencowe08}, one thus expects that a nonlinear cavity will be capable of much stronger backaction damping than a linear cavity, given the enhanced slope of the cavity response curve (c.f. Fig.~\ref{fig:Gain}).   
 
 A large backaction damping is not, however, in itself enough to ensure good cooling.  One needs that the cavity acts as a source of {\it cold} damping for the mechanical resonator.  Thus, one must also consider the effective temperature of the backaction noise, $T_{\rm BA}$. Ref.~\cite{Blencowe08} examined this quantity numerically; in contrast, our approach allows us to obtain simple analytic expressions in the interesting regime where one is near the bifurcation point and the parametric gain $G[0] \gg 1$.
 
 Our analysis is based on the unsymmetrized back action noise spectral density, defined as
\be
	S_{FF}[\omega] \equiv \int_{-\infty}^{\infty}\mathrm{d}t \langle \hat{F}(t)\hat{F}(0)\rangle.
	\label{eq:SFFUnsymm}
\ee
The symmetrized noise considered in previous sections is given by $\bS_{FF}[\omega] = (S_{FF}[\omega] +S_{FF}[-\omega] )/2$.  The frequency asymmetry of $S_{FF}[\omega]$ describes the asymmetry between emission and absorption of energy by the cavity; a standard perturbative calculation shows that it directly determines the backaction damping of the mechanical resonator \cite{ClerkRMP}:
\be
\label{eq:damping_coef}
	\gamma_{\mathrm{BA}}[\omega] = \frac{1}{2 m \hbar  \omega_m }(S_{FF}[+\omega]-S_{FF}[-\omega]).
\ee
where $m$ is the oscillator mass.  Using the linearized-Langevin approach described in previous sections, we find a particular simple form for $\gamma_{\rm BA}$ near the bifurcation, in the limit where $\omega_M / \Omega_B$ remains constant as the parametric gain $G \ra \infty$:
\be
	\label{eq:damping_nonlinear}
 		\gamma_{\rm BA} \sim   
			\frac{1}{\sqrt{3}}  \frac{A^2  \bar{n}}{ \hbar m \kappa^2 }     \frac{G[0] }{1+(\omega/\Omega_B)^2} 
	\equiv 
			\frac{1}{\sqrt{3}}  \frac{A^2  \bar{n}}{ \hbar m \kappa^2 }     G[\omega].
\ee
In the $\omega_M \ra 0$ limit, this reproduces the classical expression $\gamma_{\rm BA} \propto d \nbar / d \Delta$, while for non-zero frequency, we see that the backaction damping decays rapidly on the scale of the parametric amplification bandwidth $\Omega_B$.  The full expression (valid even for small $G$) is given in Appendix \ref{sec:appendix_dpa}.  It is instructive to compare this result for $\gamma_{\rm BA}$ against the corresponding expression for a linear cavity, in the relevant limit $\omega_M \ll \kappa$, and for an optimized detuning \cite{WilsonRae07, Marquardt07}.  As expected, one finds that the nonlinear cavity's $\gamma_{\rm BA}$ is enhanced by a factor of the parametric gain $G[\omega]$.  

As already discussed, we must also consider the effective temperature $T_{\rm BA}[\omega]$ of the backaction, a quantity which is in general frequency dependent and is defined as \cite{ClerkRMP}
\be
\label{eq:temp_eff}
	\exp \left[ -\frac{ \hbar \omega }{ k_B T_{\rm BA}[\omega] } \right] \equiv
		\frac{ S_{FF}[-\omega] }{S_{FF}[+\omega]}.
\ee
Using our linearized Langevin approach, we find a particularly simple asymptotic expression for the Bose-Einstein factor $n_{\rm BA}$ associated with $T_{\rm BA}[\omega]$ in the large-$G$ limit relevant near the bifurcation:
\be
	\label{eq:occ_number}
	1+2n_{\mathrm{BA}}[\omega]
	\equiv \coth\left( \frac{\hbar \omega}{2 k_B T_{\rm BA}[\omega] } \right)
	 \sim \frac{1+3(\omega/\kappa)^2}{\sqrt{3}(\omega/\kappa)}.
\ee
The full expression for $n_{\rm BA}$ is given in Appendix \ref{sec:appendix_dpa}.  The expression for $n_{\rm BA}[\omega]$ is remarkably similar to the corresponding expression for a linear cavity \cite{WilsonRae07, Marquardt07}.  In particular, the relevant frequency scale is $\kappa$, and {\it not} the much smaller scale set by the parametric bandwidth $\Omega_B$.  One thus finds that in the low-frequency limit $\omega_M \ll \kappa$:
\be
	T_{\rm BA}[0] \sim \frac{\kappa}{2 \sqrt{3} }.
\ee
In contrast, in the low-frequency limit, the effective backaction temperature of an optimally-driven linear cavity is $\kappa / 2$.  

Thus, the effective backaction temperature of our nonlinear cavity near the bifurcation only differs by a numerical prefactor from that of a linear cavity.  For low-frequencies, the final oscillator temperature $T_{\rm osc}$ is given by \cite{Marquardt07}:
\be
	\label{eq:T_osc}
	T_{\mathrm{osc}} = \frac{  \gamma_{\rm 0} T_0 + \gamma_{\mathrm{BA}} T_{\mathrm{BA}}   }
		{  \gamma_0+\gamma_{\mathrm{BA}}  },
\ee
Here, $\gamma_0$ is the oscillator damping resulting from its intrinsic (i.e.~non-backaction) sources of dissipation, and $T_0$ is the temperature of 
this bath.  We have thus established that for a low frequency mechanical resonator, the nonlinear cavity is a far better way to cool than the linear cavity.  One has a much greater backaction damping rate, as well as a slightly smaller effective backaction temperature.


\section{Conclusions}

In this paper, we have given a theoretical treatment of the quantum measurement properties of a driven nonlinear cavity used as a linear 
detector or amplifier.  By using the equivalence between this system near its bifurcation point and a degenerate parametric amplifier driven by a detuned pump, we were able to give a relatively simple description of the physics.  We find that quantum limited amplification is indeed possible, but only if one is able to make use of the large correlations between backaction and imprecision noises.  Such correlations are ideally suited to position detection of a mechanical system far from resonance; however, they cannot be utilized in QND qubit detection, and hence one is far from reaching the relevant quantum limit on this task.  We also examined the possibility of backaction cooling using this system, demonstrating that the nonlinearity is particularly useful in the case where one wants to cool a mechanical resonator whose frequency $\omega_M \ll \kappa$.  

\section*{Acknowledgements}

We thank K. Lehnert and R. Vijay for useful discussions.  This work was supported by NSERC, FQRNT, 
and the Canadian Institute for Advanced Research.  

\begin{appendix}

\section{}
\label{sec:appendix_dpa}
\subsection{Mapping to the Detuned DPA}
\label{sec:appendix_map}

Using Eqs.~(\ref{eq:CavitydEOM} ) and (\ref{eq:hamiltonian_dpa}), the equation of motion for the displaced cavity annihilation operator $\hd$ takes the form:
\begin{eqnarray}
   \dot{\hd} &= &(-\kappa/2 + i\tdel)\:\hd +\tlam \:\hddag - \sqrt{\kappa}  \hxi(t).
\end{eqnarray}
Introducing the canonical quadratures
\be
   \begin{pmatrix}
   \hat{x} \\
   \hat{p}
   \end{pmatrix} = \frac{1}{\sqrt{2}}
   \begin{bmatrix}
     1 & 1 \\
     -i  &  i
   \end{bmatrix} 
   \begin{pmatrix}
     \hat{d} \\
     \hat{d}^\dagger
   \end{pmatrix},
\ee
and defining $\hat{x}_{\rm in}, \hat{p}_{\rm in}$ to be the corresponding quadratures of the noise operator $\hxi$, the equations of motion take the form:
\begin{equation}
  \label{eq:a_eom_matrix}
  \frac{d}{dt} \begin{pmatrix}
    \hat{x} \\
    \hat{p}
  \end{pmatrix} = \M
  \begin{pmatrix}
    \hat{x} \\
    \hat{p}
  \end{pmatrix}-\sqrt{\kappa}
  \begin{pmatrix}
    \hat{x}_{\mathrm{in}}\\
    \hat{p}_{\mathrm{in}}.
  \end{pmatrix}
\end{equation}
Here, $\M$ is the matrix defined as
\be
  \label{eq:matrix_M1}
  \M =
  \begin{bmatrix}
    \tlam -\frac{\kappa}{2} & -\tdel \\
    \tdel  &  -(\tlam +\frac{\kappa}{2})
  \end{bmatrix} .
  \ee
  
Eq.~(\ref{eq:a_eom_matrix}) can be conveniently solved by first diagonalizing $\M$.  The only subtlety is that due to the nonzero effective drive detuning $\tdel$, $\M$ is non-Hermitian; as a result, its eigenvectors are not orthogonal to one another.  Defining $\theta$ as per Eq.~(\ref{eq:theta_def}), we let:
\begin{eqnarray}
	\mathbf{V} = 
		\begin{bmatrix}
     			\cos(\theta/2) &\sin(\theta/2) \\
     			\sin(\theta/2)  & \cos(\theta/2)
   		\end{bmatrix} 
\end{eqnarray}
denote the matrix whose columns are the eigenvectors of $\M$. One then has:
\be
	\M = -\mathbf{V} 
	\begin{bmatrix}
		\left( \chi_1[0] \right)^{-1}	&	0	\\
		0					&	\left( \chi_2[0] \right)^{-1}	
	\end{bmatrix}
	\mathbf{V}^{-1}
	\label{eq:MDiagonalized}
\ee
where the eigenvalues of $\M$ are just the inverses of the susceptibilities $\chi_1[0], \chi_2[0]$ defined in Eq.~(\ref{eqs:Chis}).

The rotation to the quadratures $\hX$ and $\hP$ introduced in Eq.~(\ref{eqs:NewQuadsX}) can now be written as
 \be
   \label{eq:a_definequad}
   \begin{pmatrix}
     \hat{X} \\
     \hat{P}
   \end{pmatrix} \equiv \T  \begin{pmatrix}
     \hat{x} \\
     \hat{p}
   \end{pmatrix}=
   \begin{bmatrix}
     \cos{\left(\theta/2\right)} &\sin{\left(\theta/2\right)} \\
     -\sin{\left(\theta/2\right)} &\cos{\left(\theta/2\right)}
   \end{bmatrix} 
   \begin{pmatrix}
     \hat{x} \\
     \hat{p}
   \end{pmatrix}.
\ee
The form of $\M$ makes it clear that $\hX$ as defined in Eq.~(\ref{eqs:NewQuadsX}) is indeed the amplified ``eigenquadrature" of the cavity:  it corresponds to the first eigenvector and eigenvalue of $\M$.  In contrast, the orthogonal quadrature $\hP$ defined in Eq.~(\ref{eqs:NewQuadsP}) {\it does not} correspond to an eigenvector of $\M$.

Finally, Fourier transforming the equations of motion Eq.~(\ref{eq:a_definequad}) using the convention:
\be
\hat{A}[\omega] \equiv \int_{-\infty}^{\infty}\mathrm{d}t \hat{A}(t)e^{-i\omega t}.
\ee
and making the above rotation, we find:
\be
 (i\omega \mathbf{1} + \T\M\T^{-1}) \begin{pmatrix}
  \hat{X}[\omega] \\
  \hat{P}[\omega]
 \end{pmatrix} = \sqrt{\kappa}
\begin{pmatrix}
  \hat{X}_{\mathrm{in}}[\omega] \\
  \hat{P}_{\mathrm{in}}[\omega]
\end{pmatrix}.
\ee
where the identity  matrix is $\mathbf{1}_{\mathrm{ij}}=\delta_{\mathrm{ij}}$.   Solving for $ \hat{X}[\omega]$ and $\hat{P}[\omega]$ directly yields Eqs.~(\ref{eqs:EOMSols}).

\subsection{Back Action Force}

Using the definition of the backaction force operator $\hF$ given in Eq.~(\ref{eq:force}) and solutions to the cavity equations of motion, Eqs.~(\ref{eqs:EOMSols}), we find: 
\be
   \label{eq:force_allorders}
   \hF[\omega] =  F_{\mathrm{x}}[\omega]\; \hat{X}_{\mathrm{in}}[\omega] + F_{\mathrm{p}}[\omega]\; \hat{P}_{\mathrm{in}}[\omega],
\ee
where
\bse
\label{eqs:FOpDecomp}
\ba
   \label{eq:define_a1a2}
   F_{\mathrm{x}}[\omega] &\equiv& - A \sqrt{2 \photonavg \kappa}\sin{\nu}\chi_1[\omega],
   	\nonumber \\
   F_{\mathrm{p}}[\omega] &\equiv&  - A \sqrt{2 \photonavg \kappa}\big[(\chi_2[\omega]-\chi_1[\omega])\sin{\nu}\tan{\theta} \nonumber \\
&&\qquad \qquad \qquad \qquad + \,\chi_2[\omega]\cos{\nu}\big],
\ea
\ese
and the angle $\nu$ is defined in Eq.~(\ref{eq:nu}).

The unsymmetrized force noise spectral density $S_{FF}[\omega]$ defined in Eq.(\ref{eq:SFFUnsymm}) can be written in terms of $\hF[\omega]$ as:
\ba
 \label{eq:SF_frequency}
   2\pi \delta(\omega+\omega')S_{FF}[\omega] & = & \langle \hF[\omega] \hF[\omega'] \rangle .
 \ea
We can thus use Eq.~(\ref{eq:force_allorders}) to calculate $S_{FF}[\omega]$ if we know the correlation functions of the input noise operators 
$\hX_{\rm in}$, $\hP_{\rm in}$.  From standard input-output theory, and our assumption that $\hxi(t)$ describes vacuum noise, one easily finds:
\bse
	\label{eqs:InputCorrelators}
\ba
	\langle \hat{X}_{\mathrm{in}}[\omega]\hat{X}_{\mathrm{in}}[\omega']\rangle& =& \pi \delta(\omega+\omega') \\
	\langle \hat{P}_{\mathrm{in}}[\omega]\hat{P}_{\mathrm{in}}[\omega']\rangle &= & \pi \delta(\omega+\omega') \\
	\langle \hat{X}_{\mathrm{in}}[\omega]\hat{P}_{\mathrm{in}}[\omega']\rangle& =& i \pi \delta(\omega+\omega')  .
\ea
\ese
Explicitly computing the symmetrized spectral density $\bar{S}_{FF}[\omega] = ( S_{FF}[\omega] + S_{FF}[-\omega] ) / 2$,  we find:
\ba
\label{eq:specFF}
	&& \bar{S}_{FF}[\omega] =   4 A^2\photonavg  \kappa \times
	\\
	&&
		\left( \frac{\kappa^2 + 6\tlam^2 + 4\omega^2 - 2\tlam^2 (\cos 2\theta + 4 \sin\theta   )}{(\kappa^2+4\omega^2)^2
			-8\tlam^2\cos^2{\theta}(\kappa^2-4\omega^2-2\tlam^2\cos^2{\theta})}
		\right).
			\nonumber 
\ea
We have used the value of the angle $\nu$ given in Eq.~(\ref{eq:nu}).  We stress that this expression only involves our initial linearization of the dynamics, and does involve any further assumption of being close to the bifurcation.   In the limit where one approaches the bifurcation 
(i.e.~ $\kappa \chi_1[0] \ra \infty$, $\theta \ra \pi/6$), one obtains the asymptotic form given in Eq.~(\ref{eq:SFFLeading}).

One can use Eq.~(\ref{eq:specFF}) to verify that $\bS_{FF}[0]$ is indeed related to the derivative of $\langle \ha \rangle$ with respect to $\Delta$ as per 
Eqs.~(\ref{eq:SFFOverlap}) and (\ref{eq:SFFApprox}).  This is easily done using $\langle \ha \rangle = \sqrt{\bar{n}} e^{i \phi_a}$, where $\bar{n}$ is given by Eq.~(\ref{eq:bif_eta}), and the phase $\phi_a$ is given by:
\begin{eqnarray}
	\tan \phi_a =  -\frac{\kappa}{2 \Delta + 4 \Lambda \bar{n}}
\end{eqnarray}
(as follows from the classical equations of motion).

Finally, we note that the above results are easily generalized to finite temperature.  The input noise correlators in Eqs.~(\ref{eqs:InputCorrelators}) and 
$\bS_{FF}[\omega]$ are simply multiplied by $(1 + \bar{n}_{\rm th})$, where $\bar{n}_{\rm th}$ is a Bose-Einstein factor evaluated at the temperature of the incident thermal noise, and the frequency of the cavity.

\subsection{Imprecision Noise}

The intensity $\hat{I}[\omega]$ of the homodyne measurement is given by Eq.~(\ref{eq:IQuads}). Using the solutions to the equation of motion in 
Eq.~(\ref{eqs:EOMSols}) we obtain  
\be
   \label{eq:i_allorders}
   \hat{I}[\omega] =  I_{\mathrm{x}}[\omega]\; \hat{X}_{\mathrm{in}} + I_{\mathrm{p}}[\omega]\; \hat{P}_{\mathrm{in}},
\ee
where
\bse
\ba
   I_{\mathrm{x}}[\omega] &=&\sqrt{\kappa} \cos{\varphi_h}(1- \kappa \chi_1[\omega])  \\
   I_{\mathrm{p}}[\omega] &=&\sqrt{\kappa} \big(\sin{\varphi_h}(1- \kappa \chi_2[\omega])  \\
&&\qquad \;- \kappa \cos{\varphi_h}\tan{\theta}(\chi_2[\omega]- \chi_1[\omega])\big). \nonumber
\ea
\ese
It is now a straightforward exercise to compute the spectral density $S_{II}[\omega]$ from Eq.~(\ref{eq:i_allorders}) and 
Eqs.~(\ref{eqs:InputCorrelators}), in complete analogy to our calculation of $S_{FF}[\omega]$.  One finds that 
this output noise completely symmetric in frequency: $S_{II}[\omega]  = S_{II}[-\omega]$, and thus $S_{II}[\omega] = \bar{S}_{II}[\omega]$.
The imprecision noise spectral density $\bS_{zz}[\omega]$ then follows using Eqs.~(\ref{eq:ImprecisionNoiseDefn}) and (\ref{eq:ChiIF}).


\subsection{Imprecision - backaction correlation}

The symmetrized imprecision-backaction noise correlator $\bS_{IF}[\omega]$ may be written:
\be
	\bar{S}_{IF}[\omega] = \frac{1}{2}\left(S_{IF}[\omega] +S_{IF}[-\omega]^*\right).
\ee
where the unsymmetrized correlator is given by:
\ba
   2 \pi \delta(\omega+\omega')S_{IF}[\omega] &=& \langle  \hat{I}[\omega] \hF[\omega'] \rangle. \nonumber \\
 \ea

We may thus calculate $\bS_{IF}[\omega]$ using Eq. (\ref{eq:force_allorders}), (\ref{eq:i_allorders}) and (\ref{eqs:InputCorrelators}); dividing
by $\chi_{IF}[\omega]$ as given in Eq.~(\ref{eq:ChiIF}) then yields the desired correlator $\bS_{zF}[\omega]$.


\subsection{Cooling}

Using Eqs.~(\ref{eqs:FOpDecomp}), on finds that the full expression for the asymmetric-in-frequency part of $S_{FF}[\omega]$ is given by:
\ba
	 &&	\frac{ S_{FF}[+\omega]-S_{FF}[-\omega]}{\omega}  = 
	 	\label{eq:FullSFFAsymm}	\\
	&&
	 \frac{  64 A^2  \bar{n} \tlam \kappa \left( \sin \theta - 1 \right) }  
	{	(\kappa^2+4\omega^2)^2 -
		8 \tlam^2 \cos^2 \theta \left(
			\kappa^2 - 4 \omega^2 - 2 \tlam^2 \cos^2 \theta
		\right)
	}. \nonumber
 \ea
This expression then directly gives the backaction damping via Eq.~(\ref{eq:damping_coef}).

Combining the above expression with Eq.~(\ref{eq:specFF}) for  $\bS_{FF}[\omega]$, we can find a general expression for $n_{\rm BA}[\omega]$, the effective temperature of the backaction expressed as a number of quanta (c.f.~Eq.(\ref{eq:occ_number})).  We have:
\ba
	1+2n_{\rm BA}[\omega] &=& \frac{		
			\kappa^2+6\tlam^2+4\omega^2-2\tlam^2(\cos{2\theta}+4\sin{\theta}) 
		}
	{8\tlam\omega(\sin{\theta}-1)}.
	\nonumber \\
	\label{eq:FullTEff}
\ea
We stress that Eqs.~(\ref{eq:FullSFFAsymm}) and (\ref{eq:FullTEff}) do not involve an assumption of being near the bifurcation.  

\subsection{Reverse Gain}
The forward gain in our system was defined in Eq.~(\ref{eq:ChiIFDefn}), which upon Fourier transforming takes the form:
\be
\hat{I}[\omega] = \chi_{IF}[\omega]\hz[\omega] .
\ee
We derived $\chi_{IF}[\omega]$ in the main text by accounting for the coupling to $\hz$ in the cavity equations of motion, resulting in 
Eq.~(\ref{eq:ChiIF}).

In general, an amplifier may also have a reverse gain $\chi_{FI}[\omega]$; this describes how signals coupled to the output operator $\hI$ could affect the average value of the backaction force operator $\hF$ \cite{ClerkRMP}.  In general, reverse gain is undesirable, as it implies that measuring the detector output (by coupling to it) can lead to enhanced backaction fluctuations.  The forms of the fundamental quantum noise inequality of Eq.~(\ref{eq:QNConstraintZ}) are also modified in the presence of reverse gain.  

To show that the reverse gain of our cavity amplifier vanishes, we make use of the equation \cite{ClerkRMP}:
\be
\label{eq:def_rev_gain}
\chi_{IF}[\omega]-\chi_{FI}[\omega]^* = - (i/\hbar)[S_{IF}[\omega]-S_{IF}[-\omega]^*].
\ee

Using the solution of the cavity equations of motion to calculate  $S_{IF}[\omega]$, and using the expression for $\chi_{IF}[\omega]$ from 
Eq.~(\ref{eq:ChiIF}) , Eq.~(\ref{eq:def_rev_gain}) directly yields that there is no reverse gain at any frequency:
\be
\chi_{FI}[\omega]=0.
\ee 

\end{appendix}

\bibliographystyle{apsrev}
\bibliography{ACTotalRefsV2}

\begin{thebibliography}{43}
\expandafter\ifx\csname natexlab\endcsname\relax\def\natexlab#1{#1}\fi
\expandafter\ifx\csname bibnamefont\endcsname\relax
  \def\bibnamefont#1{#1}\fi
\expandafter\ifx\csname bibfnamefont\endcsname\relax
  \def\bibfnamefont#1{#1}\fi
\expandafter\ifx\csname citenamefont\endcsname\relax
  \def\citenamefont#1{#1}\fi
\expandafter\ifx\csname url\endcsname\relax
  \def\url#1{\texttt{#1}}\fi
\expandafter\ifx\csname urlprefix\endcsname\relax\def\urlprefix{URL }\fi
\providecommand{\bibinfo}[2]{#2}
\providecommand{\eprint}[2][]{\url{#2}}

\bibitem[{\citenamefont{Teufel et~al.}(2009)\citenamefont{Teufel, Donner,
  Castellanos-Beltrana, Harlow, and Lehnert}}]{Lehnert09}
\bibinfo{author}{\bibfnamefont{J.~D.} \bibnamefont{Teufel}},
  \bibinfo{author}{\bibfnamefont{T.}~\bibnamefont{Donner}},
  \bibinfo{author}{\bibfnamefont{M.~A.} \bibnamefont{Castellanos-Beltrana}},
  \bibinfo{author}{\bibfnamefont{J.~W.} \bibnamefont{Harlow}},
  \bibnamefont{and} \bibinfo{author}{\bibfnamefont{K.~W.}
  \bibnamefont{Lehnert}}, \bibinfo{journal}{Nature Nanotech.}
  \textbf{\bibinfo{volume}{4}}, \bibinfo{pages}{820} (\bibinfo{year}{2009}).

\bibitem[{\citenamefont{Hertzberg et~al.}(2010)\citenamefont{Hertzberg,
  Rocheleau, Ndukum, Savva, Clerk, and Schwab}}]{Schwab09}
\bibinfo{author}{\bibfnamefont{J.~B.} \bibnamefont{Hertzberg}},
  \bibinfo{author}{\bibfnamefont{T.}~\bibnamefont{Rocheleau}},
  \bibinfo{author}{\bibfnamefont{T.}~\bibnamefont{Ndukum}},
  \bibinfo{author}{\bibfnamefont{M.}~\bibnamefont{Savva}},
  \bibinfo{author}{\bibfnamefont{A.~A.} \bibnamefont{Clerk}}, \bibnamefont{and}
  \bibinfo{author}{\bibfnamefont{K.~C.} \bibnamefont{Schwab}},
  \bibinfo{journal}{Nature Phys.} \textbf{\bibinfo{volume}{6}},
  \bibinfo{pages}{213} (\bibinfo{year}{2010}).

\bibitem[{\citenamefont{Schuster et~al.}(2005)\citenamefont{Schuster, Wallraff,
  Blais, Frunzio, Huang, Majer, Girvin, and Schoelkopf}}]{Schuster05}
\bibinfo{author}{\bibfnamefont{D.~I.} \bibnamefont{Schuster}},
  \bibinfo{author}{\bibfnamefont{A.}~\bibnamefont{Wallraff}},
  \bibinfo{author}{\bibfnamefont{A.}~\bibnamefont{Blais}},
  \bibinfo{author}{\bibfnamefont{L.}~\bibnamefont{Frunzio}},
  \bibinfo{author}{\bibfnamefont{R.-S.} \bibnamefont{Huang}},
  \bibinfo{author}{\bibfnamefont{J.}~\bibnamefont{Majer}},
  \bibinfo{author}{\bibfnamefont{S.~M.} \bibnamefont{Girvin}},
  \bibnamefont{and} \bibinfo{author}{\bibfnamefont{R.~J.}
  \bibnamefont{Schoelkopf}}, \bibinfo{journal}{Phys. Rev. Lett.}
  \textbf{\bibinfo{volume}{94}}, \bibinfo{pages}{123602}
  (\bibinfo{year}{2005}).

\bibitem[{\citenamefont{Majer et~al.}(2007)\citenamefont{Majer, Chow, Gambetta,
  Koch, Johnson, Schreier, Frunzio, Schuster, Houck, Wallraff
  et~al.}}]{Majer07}
\bibinfo{author}{\bibfnamefont{J.}~\bibnamefont{Majer}},
  \bibinfo{author}{\bibfnamefont{J.~M.} \bibnamefont{Chow}},
  \bibinfo{author}{\bibfnamefont{J.~M.} \bibnamefont{Gambetta}},
  \bibinfo{author}{\bibfnamefont{J.}~\bibnamefont{Koch}},
  \bibinfo{author}{\bibfnamefont{B.~R.} \bibnamefont{Johnson}},
  \bibinfo{author}{\bibfnamefont{J.~A.} \bibnamefont{Schreier}},
  \bibinfo{author}{\bibfnamefont{L.}~\bibnamefont{Frunzio}},
  \bibinfo{author}{\bibfnamefont{D.~I.} \bibnamefont{Schuster}},
  \bibinfo{author}{\bibfnamefont{A.~A.} \bibnamefont{Houck}},
  \bibinfo{author}{\bibfnamefont{A.}~\bibnamefont{Wallraff}},
  \bibnamefont{et~al.}, \bibinfo{journal}{Nature}
  \textbf{\bibinfo{volume}{449}}, \bibinfo{pages}{443} (\bibinfo{year}{2007}).

\bibitem[{\citenamefont{Clerk et~al.}(2010)\citenamefont{Clerk, Devoret,
  Girvin, Marquardt, and Schoelkopf}}]{ClerkRMP}
\bibinfo{author}{\bibfnamefont{A.~A.} \bibnamefont{Clerk}},
  \bibinfo{author}{\bibfnamefont{M.~H.} \bibnamefont{Devoret}},
  \bibinfo{author}{\bibfnamefont{S.~M.} \bibnamefont{Girvin}},
  \bibinfo{author}{\bibfnamefont{F.}~\bibnamefont{Marquardt}},
  \bibnamefont{and} \bibinfo{author}{\bibfnamefont{R.~J.}
  \bibnamefont{Schoelkopf}}, \bibinfo{journal}{Rev. Mod. Phys.}
  \textbf{\bibinfo{volume}{82}}, \bibinfo{pages}{1155} (\bibinfo{year}{2010}).

\bibitem[{\citenamefont{Yurke}(1987)}]{Yurke87}
\bibinfo{author}{\bibfnamefont{B.}~\bibnamefont{Yurke}}, \bibinfo{journal}{J.
  Opt. Soc. Am. B} \textbf{\bibinfo{volume}{4}}, \bibinfo{pages}{1551}
  (\bibinfo{year}{1987}).

\bibitem[{\citenamefont{Yurke et~al.}(1989)\citenamefont{Yurke, Corruccini,
  Kaminsky, Rupp, Smith, Silver, Simon, and Whittaker}}]{Yurke89}
\bibinfo{author}{\bibfnamefont{B.}~\bibnamefont{Yurke}},
  \bibinfo{author}{\bibfnamefont{L.~R.} \bibnamefont{Corruccini}},
  \bibinfo{author}{\bibfnamefont{P.~G.} \bibnamefont{Kaminsky}},
  \bibinfo{author}{\bibfnamefont{L.~W.} \bibnamefont{Rupp}},
  \bibinfo{author}{\bibfnamefont{A.~D.} \bibnamefont{Smith}},
  \bibinfo{author}{\bibfnamefont{A.~H.} \bibnamefont{Silver}},
  \bibinfo{author}{\bibfnamefont{R.~W.} \bibnamefont{Simon}}, \bibnamefont{and}
  \bibinfo{author}{\bibfnamefont{E.~A.} \bibnamefont{Whittaker}},
  \bibinfo{journal}{Phys. Rev. A} \textbf{\bibinfo{volume}{39}},
  \bibinfo{pages}{2519} (\bibinfo{year}{1989}).

\bibitem[{\citenamefont{Nation et~al.}(2008)\citenamefont{Nation, Blencowe, and
  Buks}}]{Blencowe08}
\bibinfo{author}{\bibfnamefont{P.~D.} \bibnamefont{Nation}},
  \bibinfo{author}{\bibfnamefont{M.~P.} \bibnamefont{Blencowe}},
  \bibnamefont{and} \bibinfo{author}{\bibfnamefont{E.}~\bibnamefont{Buks}},
  \bibinfo{journal}{Phys. Rev. B} \textbf{\bibinfo{volume}{78}},
  \bibinfo{pages}{104516} (\bibinfo{year}{2008}).

\bibitem[{\citenamefont{Movshovich et~al.}(1990)\citenamefont{Movshovich,
  Yurke, Kaminsky, Smith, Silver, Simon, and Schneider}}]{Yurke90}
\bibinfo{author}{\bibfnamefont{R.}~\bibnamefont{Movshovich}},
  \bibinfo{author}{\bibfnamefont{B.}~\bibnamefont{Yurke}},
  \bibinfo{author}{\bibfnamefont{P.~G.} \bibnamefont{Kaminsky}},
  \bibinfo{author}{\bibfnamefont{A.~D.} \bibnamefont{Smith}},
  \bibinfo{author}{\bibfnamefont{A.~H.} \bibnamefont{Silver}},
  \bibinfo{author}{\bibfnamefont{R.~W.} \bibnamefont{Simon}}, \bibnamefont{and}
  \bibinfo{author}{\bibfnamefont{M.~V.} \bibnamefont{Schneider}},
  \bibinfo{journal}{Phys. Rev. Lett.} \textbf{\bibinfo{volume}{65}},
  \bibinfo{pages}{1419} (\bibinfo{year}{1990}).

\bibitem[{\citenamefont{Castellanos-Beltran
  et~al.}(2008)\citenamefont{Castellanos-Beltran, Irwin, Hilton, Vale, and
  Lehnert}}]{Lehnert08}
\bibinfo{author}{\bibfnamefont{M.~A.} \bibnamefont{Castellanos-Beltran}},
  \bibinfo{author}{\bibfnamefont{K.~D.} \bibnamefont{Irwin}},
  \bibinfo{author}{\bibfnamefont{G.~C.} \bibnamefont{Hilton}},
  \bibinfo{author}{\bibfnamefont{L.~R.} \bibnamefont{Vale}}, \bibnamefont{and}
  \bibinfo{author}{\bibfnamefont{K.~W.} \bibnamefont{Lehnert}},
  \bibinfo{journal}{Nature Phys.} \textbf{\bibinfo{volume}{4}},
  \bibinfo{pages}{929} (\bibinfo{year}{2008}).

\bibitem[{\citenamefont{Yamamoto et~al.}(2008)\citenamefont{Yamamoto, Inomata,
  Watanabe, Matsuba, Miyazaki, Oliver, Nakamura, and Tsai}}]{Nakamura08}
\bibinfo{author}{\bibfnamefont{T.}~\bibnamefont{Yamamoto}},
  \bibinfo{author}{\bibfnamefont{K.}~\bibnamefont{Inomata}},
  \bibinfo{author}{\bibfnamefont{M.}~\bibnamefont{Watanabe}},
  \bibinfo{author}{\bibfnamefont{K.}~\bibnamefont{Matsuba}},
  \bibinfo{author}{\bibfnamefont{T.}~\bibnamefont{Miyazaki}},
  \bibinfo{author}{\bibfnamefont{W.~D.} \bibnamefont{Oliver}},
  \bibinfo{author}{\bibfnamefont{Y.}~\bibnamefont{Nakamura}}, \bibnamefont{and}
  \bibinfo{author}{\bibfnamefont{J.~S.} \bibnamefont{Tsai}},
  \bibinfo{journal}{Appl. Phys. Lett.} \textbf{\bibinfo{volume}{93}},
  \bibinfo{pages}{042510} (\bibinfo{year}{2008}).

\bibitem[{\citenamefont{Yurke and Buks}(2006)}]{Yurke06}
\bibinfo{author}{\bibfnamefont{B.}~\bibnamefont{Yurke}} \bibnamefont{and}
  \bibinfo{author}{\bibfnamefont{E.}~\bibnamefont{Buks}}, \bibinfo{journal}{J.
  Lightwave. Tech.} \textbf{\bibinfo{volume}{24}}, \bibinfo{pages}{5054}
  (\bibinfo{year}{2006}).

\bibitem[{\citenamefont{Babourina-Brooks
  et~al.}(2008)\citenamefont{Babourina-Brooks, Doherty, and
  Milburn}}]{Milburn08}
\bibinfo{author}{\bibfnamefont{E.}~\bibnamefont{Babourina-Brooks}},
  \bibinfo{author}{\bibfnamefont{A.}~\bibnamefont{Doherty}}, \bibnamefont{and}
  \bibinfo{author}{\bibfnamefont{G.}~\bibnamefont{Milburn}},
  \bibinfo{journal}{New J. Phys.} \textbf{\bibinfo{volume}{10}},
  \bibinfo{pages}{105020} (\bibinfo{year}{2008}).

\bibitem[{\citenamefont{Siddiqi et~al.}(2004)\citenamefont{Siddiqi, Vijay,
  Pierre, Wilson, Metcalfe, Rigetti, Frunzio, and Devoret}}]{Siddiqi04}
\bibinfo{author}{\bibfnamefont{I.}~\bibnamefont{Siddiqi}},
  \bibinfo{author}{\bibfnamefont{R.}~\bibnamefont{Vijay}},
  \bibinfo{author}{\bibfnamefont{F.}~\bibnamefont{Pierre}},
  \bibinfo{author}{\bibfnamefont{C.~M.} \bibnamefont{Wilson}},
  \bibinfo{author}{\bibfnamefont{M.}~\bibnamefont{Metcalfe}},
  \bibinfo{author}{\bibfnamefont{C.}~\bibnamefont{Rigetti}},
  \bibinfo{author}{\bibfnamefont{L.}~\bibnamefont{Frunzio}}, \bibnamefont{and}
  \bibinfo{author}{\bibfnamefont{M.~H.} \bibnamefont{Devoret}},
  \bibinfo{journal}{Phys. Rev. Lett.} \textbf{\bibinfo{volume}{93}},
  \bibinfo{pages}{207002} (\bibinfo{year}{2004}).

\bibitem[{\citenamefont{Lupa\c{s}cu et~al.}(2006)\citenamefont{Lupa\c{s}cu,
  Driessen, Roschier, Harmans, and Mooij}}]{Lupascu06}
\bibinfo{author}{\bibfnamefont{A.}~\bibnamefont{Lupa\c{s}cu}},
  \bibinfo{author}{\bibfnamefont{E.~F.~C.} \bibnamefont{Driessen}},
  \bibinfo{author}{\bibfnamefont{L.}~\bibnamefont{Roschier}},
  \bibinfo{author}{\bibfnamefont{C.~J. P.~M.} \bibnamefont{Harmans}},
  \bibnamefont{and} \bibinfo{author}{\bibfnamefont{J.~E.} \bibnamefont{Mooij}},
  \bibinfo{journal}{Phys. Rev. Lett.} \textbf{\bibinfo{volume}{96}},
  \bibinfo{pages}{127003} (\bibinfo{year}{2006}).

\bibitem[{\citenamefont{Serban et~al.}(2010)\citenamefont{Serban, Dykman, and
  Wilhelm}}]{Wilhelm10}
\bibinfo{author}{\bibfnamefont{I.}~\bibnamefont{Serban}},
  \bibinfo{author}{\bibfnamefont{M.~I.} \bibnamefont{Dykman}},
  \bibnamefont{and} \bibinfo{author}{\bibfnamefont{F.~K.}
  \bibnamefont{Wilhelm}}, \bibinfo{journal}{Phys. Rev. A}
  \textbf{\bibinfo{volume}{81}}, \bibinfo{pages}{022305}
  (\bibinfo{year}{2010}).

\bibitem[{\citenamefont{Braginsky and Khalili}(1992)}]{Braginsky92}
\bibinfo{author}{\bibfnamefont{V.~B.} \bibnamefont{Braginsky}}
  \bibnamefont{and} \bibinfo{author}{\bibfnamefont{F.~Y.}
  \bibnamefont{Khalili}}, \emph{\bibinfo{title}{Quantum Measurement}}
  (\bibinfo{publisher}{Cambridge University Press, Cambridge},
  \bibinfo{year}{1992}).

\bibitem[{\citenamefont{Hartridge et~al.}(2010)\citenamefont{Hartridge, Vijay,
  Slichter, Clarke, and Siddiqi}}]{Vijay10}
\bibinfo{author}{\bibfnamefont{M.}~\bibnamefont{Hartridge}},
  \bibinfo{author}{\bibfnamefont{R.}~\bibnamefont{Vijay}},
  \bibinfo{author}{\bibfnamefont{D.~H.} \bibnamefont{Slichter}},
  \bibinfo{author}{\bibfnamefont{J.}~\bibnamefont{Clarke}}, \bibnamefont{and}
  \bibinfo{author}{\bibfnamefont{I.}~\bibnamefont{Siddiqi}},
  \bibinfo{journal}{arXiv:1003.2466}  (\bibinfo{year}{2010}).

\bibitem[{\citenamefont{Vijay et~al.}(2010)\citenamefont{Vijay, Slichter, and
  Siddiqi}}]{Vijay10b}
\bibinfo{author}{\bibfnamefont{R.}~\bibnamefont{Vijay}},
  \bibinfo{author}{\bibfnamefont{D.~H.} \bibnamefont{Slichter}},
  \bibnamefont{and} \bibinfo{author}{\bibfnamefont{I.}~\bibnamefont{Siddiqi}},
  \bibinfo{journal}{arXiv:1009.2969}  (\bibinfo{year}{2010}).

\bibitem[{\citenamefont{Ong et~al.}(2010)\citenamefont{Ong, Boissonneault,
  Mallet, Palacios-Laloy, Dewes, Doherty, Blais, Bertet, Vion, and
  Esteve}}]{Esteve10}
\bibinfo{author}{\bibfnamefont{F.~R.} \bibnamefont{Ong}},
  \bibinfo{author}{\bibfnamefont{M.}~\bibnamefont{Boissonneault}},
  \bibinfo{author}{\bibfnamefont{F.}~\bibnamefont{Mallet}},
  \bibinfo{author}{\bibfnamefont{A.}~\bibnamefont{Palacios-Laloy}},
  \bibinfo{author}{\bibfnamefont{A.}~\bibnamefont{Dewes}},
  \bibinfo{author}{\bibfnamefont{A.~C.} \bibnamefont{Doherty}},
  \bibinfo{author}{\bibfnamefont{A.}~\bibnamefont{Blais}},
  \bibinfo{author}{\bibfnamefont{P.}~\bibnamefont{Bertet}},
  \bibinfo{author}{\bibfnamefont{D.}~\bibnamefont{Vion}}, \bibnamefont{and}
  \bibinfo{author}{\bibfnamefont{D.}~\bibnamefont{Esteve}},
  \bibinfo{journal}{arXiv:1010.6248}  (\bibinfo{year}{2010}).

\bibitem[{\citenamefont{Carmichael et~al.}(1984)\citenamefont{Carmichael,
  Milburn, and Walls}}]{Carmichael84}
\bibinfo{author}{\bibfnamefont{H.~J.} \bibnamefont{Carmichael}},
  \bibinfo{author}{\bibfnamefont{G.~J.} \bibnamefont{Milburn}},
  \bibnamefont{and} \bibinfo{author}{\bibfnamefont{D.~F.} \bibnamefont{Walls}},
  \bibinfo{journal}{J. Phys. A} \textbf{\bibinfo{volume}{17}},
  \bibinfo{pages}{469} (\bibinfo{year}{1984}).

\bibitem[{\citenamefont{Dykman}(1978)}]{Dykman78}
\bibinfo{author}{\bibfnamefont{M.~I.} \bibnamefont{Dykman}},
  \bibinfo{journal}{Sov. Phys. Solid State} \textbf{\bibinfo{volume}{20}},
  \bibinfo{pages}{1306} (\bibinfo{year}{1978}).

\bibitem[{\citenamefont{Gardiner and Collett}(1985)}]{Gardiner85}
\bibinfo{author}{\bibfnamefont{C.~W.} \bibnamefont{Gardiner}} \bibnamefont{and}
  \bibinfo{author}{\bibfnamefont{M.~J.} \bibnamefont{Collett}},
  \bibinfo{journal}{Phys. Rev. A} \textbf{\bibinfo{volume}{31}},
  \bibinfo{pages}{3761} (\bibinfo{year}{1985}).

\bibitem[{\citenamefont{Gardiner and Zoller}(2000)}]{Gardiner00}
\bibinfo{author}{\bibfnamefont{C.~W.} \bibnamefont{Gardiner}} \bibnamefont{and}
  \bibinfo{author}{\bibfnamefont{P.}~\bibnamefont{Zoller}},
  \emph{\bibinfo{title}{Quantum Noise}} (\bibinfo{publisher}{Springer, Berlin},
  \bibinfo{year}{2000}).

\bibitem[{\citenamefont{Haus and Mullen}(1962)}]{Haus62}
\bibinfo{author}{\bibfnamefont{H.~A.} \bibnamefont{Haus}} \bibnamefont{and}
  \bibinfo{author}{\bibfnamefont{J.~A.} \bibnamefont{Mullen}},
  \bibinfo{journal}{Phys. Rev.} \textbf{\bibinfo{volume}{128}},
  \bibinfo{pages}{2407} (\bibinfo{year}{1962}).

\bibitem[{\citenamefont{Caves}(1982)}]{Caves82}
\bibinfo{author}{\bibfnamefont{C.~M.} \bibnamefont{Caves}},
  \bibinfo{journal}{Phys. Rev. D} \textbf{\bibinfo{volume}{26}},
  \bibinfo{pages}{1817} (\bibinfo{year}{1982}).

\bibitem[{\citenamefont{Devoret and Schoelkopf}(2000)}]{Devoret00}
\bibinfo{author}{\bibfnamefont{M.~H.} \bibnamefont{Devoret}} \bibnamefont{and}
  \bibinfo{author}{\bibfnamefont{R.~J.} \bibnamefont{Schoelkopf}},
  \bibinfo{journal}{Nature (London)} \textbf{\bibinfo{volume}{406}},
  \bibinfo{pages}{1039} (\bibinfo{year}{2000}).

\bibitem[{\citenamefont{Clerk et~al.}(2003)\citenamefont{Clerk, Girvin, and
  Stone}}]{Clerk03}
\bibinfo{author}{\bibfnamefont{A.~A.} \bibnamefont{Clerk}},
  \bibinfo{author}{\bibfnamefont{S.~M.} \bibnamefont{Girvin}},
  \bibnamefont{and} \bibinfo{author}{\bibfnamefont{A.~D.} \bibnamefont{Stone}},
  \bibinfo{journal}{Phys. Rev. B} \textbf{\bibinfo{volume}{67}},
  \bibinfo{pages}{165324} (\bibinfo{year}{2003}).

\bibitem[{\citenamefont{Dykman et~al.}(1994)\citenamefont{Dykman, Luchinsky,
  Mannella, McClintock, Stein, and Stocks}}]{Dykman94}
\bibinfo{author}{\bibfnamefont{M.~I.} \bibnamefont{Dykman}},
  \bibinfo{author}{\bibfnamefont{D.~G.} \bibnamefont{Luchinsky}},
  \bibinfo{author}{\bibfnamefont{R.}~\bibnamefont{Mannella}},
  \bibinfo{author}{\bibfnamefont{P.~V.~E.} \bibnamefont{McClintock}},
  \bibinfo{author}{\bibfnamefont{N.~D.} \bibnamefont{Stein}}, \bibnamefont{and}
  \bibinfo{author}{\bibfnamefont{N.~G.} \bibnamefont{Stocks}},
  \bibinfo{journal}{Phys. Rev. E} \textbf{\bibinfo{volume}{49}},
  \bibinfo{pages}{1198} (\bibinfo{year}{1994}).

\bibitem[{\citenamefont{Blais et~al.}(2004)\citenamefont{Blais, Huang,
  Wallraff, Girvin, and Schoelkopf}}]{Blais04}
\bibinfo{author}{\bibfnamefont{A.}~\bibnamefont{Blais}},
  \bibinfo{author}{\bibfnamefont{R.-S.} \bibnamefont{Huang}},
  \bibinfo{author}{\bibfnamefont{A.}~\bibnamefont{Wallraff}},
  \bibinfo{author}{\bibfnamefont{S.~M.} \bibnamefont{Girvin}},
  \bibnamefont{and} \bibinfo{author}{\bibfnamefont{R.~J.}
  \bibnamefont{Schoelkopf}}, \bibinfo{journal}{Phys. Rev. A}
  \textbf{\bibinfo{volume}{69}}, \bibinfo{pages}{062320}
  (\bibinfo{year}{2004}).

\bibitem[{\citenamefont{Gambetta et~al.}(2006)\citenamefont{Gambetta, Blais,
  Schuster, Wallraff, Frunzio, Majer, Devoret, Girvin, and
  Schoelkopf}}]{Gambetta06}
\bibinfo{author}{\bibfnamefont{J.}~\bibnamefont{Gambetta}},
  \bibinfo{author}{\bibfnamefont{A.}~\bibnamefont{Blais}},
  \bibinfo{author}{\bibfnamefont{D.~I.} \bibnamefont{Schuster}},
  \bibinfo{author}{\bibfnamefont{A.}~\bibnamefont{Wallraff}},
  \bibinfo{author}{\bibfnamefont{L.}~\bibnamefont{Frunzio}},
  \bibinfo{author}{\bibfnamefont{J.}~\bibnamefont{Majer}},
  \bibinfo{author}{\bibfnamefont{M.~H.} \bibnamefont{Devoret}},
  \bibinfo{author}{\bibfnamefont{S.~M.} \bibnamefont{Girvin}},
  \bibnamefont{and} \bibinfo{author}{\bibfnamefont{R.~J.}
  \bibnamefont{Schoelkopf}}, \bibinfo{journal}{Phys. Rev. A}
  \textbf{\bibinfo{volume}{74}}, \bibinfo{pages}{042318}
  (\bibinfo{year}{2006}).

\bibitem[{\citenamefont{Milburn and Walls}(1981)}]{Milburn81}
\bibinfo{author}{\bibfnamefont{G.~J.} \bibnamefont{Milburn}} \bibnamefont{and}
  \bibinfo{author}{\bibfnamefont{D.~F.} \bibnamefont{Walls}},
  \bibinfo{journal}{Opt. Comm.} \textbf{\bibinfo{volume}{39}},
  \bibinfo{pages}{401} (\bibinfo{year}{1981}).

\bibitem[{\citenamefont{Corbitt and Mavalvala}(2004)}]{Corbitt04}
\bibinfo{author}{\bibfnamefont{T.}~\bibnamefont{Corbitt}} \bibnamefont{and}
  \bibinfo{author}{\bibfnamefont{N.}~\bibnamefont{Mavalvala}},
  \bibinfo{journal}{J. Opt. B} \textbf{\bibinfo{volume}{6}},
  \bibinfo{pages}{S675} (\bibinfo{year}{2004}).

\bibitem[{\citenamefont{Buonanno and Chen}(2001)}]{Chen01}
\bibinfo{author}{\bibfnamefont{A.}~\bibnamefont{Buonanno}} \bibnamefont{and}
  \bibinfo{author}{\bibfnamefont{Y.}~\bibnamefont{Chen}},
  \bibinfo{journal}{Phys. Rev. D} \textbf{\bibinfo{volume}{64}},
  \bibinfo{pages}{042006} (\bibinfo{year}{2001}).

\bibitem[{\citenamefont{Marquardt and Girvin}(2009)}]{Marquardt09}
\bibinfo{author}{\bibfnamefont{F.}~\bibnamefont{Marquardt}} \bibnamefont{and}
  \bibinfo{author}{\bibfnamefont{S.~M.} \bibnamefont{Girvin}},
  \bibinfo{journal}{Physics} \textbf{\bibinfo{volume}{2}}, \bibinfo{pages}{40}
  (\bibinfo{year}{2009}).

\bibitem[{\citenamefont{Wilson-Rae et~al.}(2007)\citenamefont{Wilson-Rae,
  Nooshi, Zwerger, and Kippenberg}}]{WilsonRae07}
\bibinfo{author}{\bibfnamefont{I.}~\bibnamefont{Wilson-Rae}},
  \bibinfo{author}{\bibfnamefont{N.}~\bibnamefont{Nooshi}},
  \bibinfo{author}{\bibfnamefont{W.}~\bibnamefont{Zwerger}}, \bibnamefont{and}
  \bibinfo{author}{\bibfnamefont{T.~J.} \bibnamefont{Kippenberg}},
  \bibinfo{journal}{Phys. Rev. Lett.} \textbf{\bibinfo{volume}{99}},
  \bibinfo{pages}{093901} (\bibinfo{year}{2007}).

\bibitem[{\citenamefont{Marquardt et~al.}(2007)\citenamefont{Marquardt, Chen,
  Clerk, and Girvin}}]{Marquardt07}
\bibinfo{author}{\bibfnamefont{F.}~\bibnamefont{Marquardt}},
  \bibinfo{author}{\bibfnamefont{J.~P.} \bibnamefont{Chen}},
  \bibinfo{author}{\bibfnamefont{A.~A.} \bibnamefont{Clerk}}, \bibnamefont{and}
  \bibinfo{author}{\bibfnamefont{S.~M.} \bibnamefont{Girvin}},
  \bibinfo{journal}{Phys. Rev. Let.} \textbf{\bibinfo{volume}{99}},
  \bibinfo{pages}{093902} (\bibinfo{year}{2007}).

\bibitem[{\citenamefont{Teufel et~al.}(2008)\citenamefont{Teufel, Harlow,
  Regal, and Lehnert}}]{Teufel08}
\bibinfo{author}{\bibfnamefont{J.~D.} \bibnamefont{Teufel}},
  \bibinfo{author}{\bibfnamefont{J.~W.} \bibnamefont{Harlow}},
  \bibinfo{author}{\bibfnamefont{C.~A.} \bibnamefont{Regal}}, \bibnamefont{and}
  \bibinfo{author}{\bibfnamefont{K.~W.} \bibnamefont{Lehnert}},
  \bibinfo{journal}{Phys. Rev. Lett.} \textbf{\bibinfo{volume}{101}},
  \bibinfo{pages}{197203} (\bibinfo{year}{2008}).

\bibitem[{\citenamefont{Rocheleau et~al.}(2010)\citenamefont{Rocheleau, Ndukum,
  Macklin, Hertzberg, Clerk, and Schwab}}]{Schwab10}
\bibinfo{author}{\bibfnamefont{T.}~\bibnamefont{Rocheleau}},
  \bibinfo{author}{\bibfnamefont{T.}~\bibnamefont{Ndukum}},
  \bibinfo{author}{\bibfnamefont{C.}~\bibnamefont{Macklin}},
  \bibinfo{author}{\bibfnamefont{J.~B.} \bibnamefont{Hertzberg}},
  \bibinfo{author}{\bibfnamefont{A.}~\bibnamefont{Clerk}}, \bibnamefont{and}
  \bibinfo{author}{\bibfnamefont{K.}~\bibnamefont{Schwab}},
  \bibinfo{journal}{Nature} \textbf{\bibinfo{volume}{463}}, \bibinfo{pages}{72}
  (\bibinfo{year}{2010}).

\bibitem[{\citenamefont{Groblacher et~al.}(2009)\citenamefont{Groblacher,
  Hertzberg, Vanner, Cole, Gigan, Schwab, and Aspelmeyer}}]{Aspelmeyer09}
\bibinfo{author}{\bibfnamefont{S.}~\bibnamefont{Groblacher}},
  \bibinfo{author}{\bibfnamefont{J.~B.} \bibnamefont{Hertzberg}},
  \bibinfo{author}{\bibfnamefont{M.~R.} \bibnamefont{Vanner}},
  \bibinfo{author}{\bibfnamefont{G.~D.} \bibnamefont{Cole}},
  \bibinfo{author}{\bibfnamefont{S.}~\bibnamefont{Gigan}},
  \bibinfo{author}{\bibfnamefont{K.~C.} \bibnamefont{Schwab}},
  \bibnamefont{and}
  \bibinfo{author}{\bibfnamefont{M.}~\bibnamefont{Aspelmeyer}},
  \bibinfo{journal}{Nature Phys.} \textbf{\bibinfo{volume}{5}},
  \bibinfo{pages}{485} (\bibinfo{year}{2009}).

\bibitem[{\citenamefont{Park and Wang}(2009)}]{Wang09}
\bibinfo{author}{\bibfnamefont{Y.-S.} \bibnamefont{Park}} \bibnamefont{and}
  \bibinfo{author}{\bibfnamefont{H.}~\bibnamefont{Wang}},
  \bibinfo{journal}{Nature Phys.} \textbf{\bibinfo{volume}{5}},
  \bibinfo{pages}{489} (\bibinfo{year}{2009}).

\bibitem[{\citenamefont{Schliesser et~al.}(2009)\citenamefont{Schliesser,
  Arcizet, Riviere, Anetsberger, and Kippenberg}}]{Kippenberg09}
\bibinfo{author}{\bibfnamefont{A.}~\bibnamefont{Schliesser}},
  \bibinfo{author}{\bibfnamefont{O.}~\bibnamefont{Arcizet}},
  \bibinfo{author}{\bibfnamefont{R.}~\bibnamefont{Riviere}},
  \bibinfo{author}{\bibfnamefont{G.}~\bibnamefont{Anetsberger}},
  \bibnamefont{and} \bibinfo{author}{\bibfnamefont{T.~J.}
  \bibnamefont{Kippenberg}}, \bibinfo{journal}{Nature Phys.}
  \textbf{\bibinfo{volume}{5}}, \bibinfo{pages}{509} (\bibinfo{year}{2009}).

\bibitem[{\citenamefont{{H\"ohberger}-Metzger and
  Karrai}(2004)}]{2004_12_HoehbergerKarrai_CoolingMicroleverNature}
\bibinfo{author}{\bibfnamefont{C.}~\bibnamefont{{H\"ohberger}-Metzger}}
  \bibnamefont{and} \bibinfo{author}{\bibfnamefont{K.}~\bibnamefont{Karrai}},
  \bibinfo{journal}{Nature} \textbf{\bibinfo{volume}{432}},
  \bibinfo{pages}{1002} (\bibinfo{year}{2004}).

\end{thebibliography}
\end{document}